\documentclass[usenatbib, useAMS]{mnras}
\usepackage{natbib}
\usepackage{graphicx}
\usepackage{rotating}
\usepackage{verbatim}
\usepackage{amssymb}
\usepackage{amsmath}
\usepackage{gensymb}
\usepackage{pdflscape}
\usepackage{dcolumn}
\newcolumntype{d}[1]{D{.}{.}{#1}}

\begin{document}

\title[The Multi --Wavelength Tully-Fisher relation]
{The Multi-Wavelength Tully-Fisher relation with spatially resolved H{\sc i} kinematics}
\author[Ponomareva et al.]
{Anastasia A. Ponomareva$^{1,2}$\thanks{E-mail:
ponomareva@astro.rug.nl}, Marc A. W. Verheijen$^{2,3}$ , Reynier F. Peletier$^2$ \newauthor
and Albert Bosma${^4}$\vspace*{0.2cm}\\
  $^1$Research School of Astronomy \& Astrophysics, Australian National University, Canberra, ACT 2611,
  Australia\\
  $^2$Kapteyn Astronomical Institute, University of Groningen, Postbus 800, 
  NL-9700 AV Groningen, The Netherlands\\
  $^3$Adjunct Faculty, National Centre for Radio Astrophysics, TIFR, Ganeshkhind, Pune 411007, India \\
  $^4$Aix Marseille Univ, CNRS, LAM, Laboratoire d'Astrophysique de Marseille, UMR 7326, F-13388, Marseille, France\\}

\pagerange{\pageref{firstpage}--\pageref{lastpage}}
\pubyear{2017}

\maketitle
\label{firstpage}
\begin{abstract}
 In this paper we investigate the statistical properties of the Tully-Fisher relation for a sample 
 of 32 galaxies with measured distances from the Cepheid period--luminosity relation and/or TRGB stars. 
 We take advantage of panchromatic photometry in 12 bands (from FUV to 4.5 $\mu$m) and 
 of spatially resolved H{\sc i} kinematics. We use these data together with three kinematic measures ($W^{i}_{50}$, $V_{max}$ and $V_{flat}$)
  extracted from the global H{\sc i} profiles or H{\sc i} rotation curves, so as to construct 36 correlations 
  allowing us to select the one with the least scatter. We introduce a tightness parameter $\sigma_{\perp}$ of the TFr,
  in order to obtain a slope--independent measure of the goodness of fit. We find that the tightest correlation occurs 
  when we select the 3.6 $\mu$m photometric band together with the $V_{flat}$ parameter extracted from the H{\sc i} rotation curve.


\end{abstract}


\defcitealias{TC12}{TC12}
\defcitealias{PaperI}{Paper I}

\section{Introduction}
The Tully-Fisher relation (TFr) is a power--law correlation between the 
luminosity and the rotation velocity of late-type galaxies \citep{tf77}. It was 
empirically established as a powerful tool to measure distances to galaxies 
independently from their redshift.
 Knowing only a galaxy's rotational velocity 
from the width of its neutral hydrogen (H{\sc i}) line profile, one can recover 
the distance modulus to this galaxy by inferring the total intrinsic luminosity from a 
calibrator sample. Thus, to obtain accurate distances, a number of studies of 
the statistical properties of the TFr were done in the past, aiming to reduce as 
much as possible the observed scatter in the relation, e.g. 
\href{http://www.ipnl.in2p3.fr/projet/cosmicflows/}{the Cosmic Flows programme} \citep{cosflows1,cosflows2, TC12}.  

Understanding the origin of the TFr is one of the main challenges for theories 
of galaxy formation and evolution. From a theoretical point of view, a perfect correlation 
between the intrinsic luminosity and rotational velocity of a galaxy is currently explained 
as a relation between the hosting dark matter halo and its 
baryonic content, assuming a direct link between luminosity and baryonic mass.
The detailed statistical properties of the TFr 
provide important constraints to semi-analytical models and numerical simulations of galaxy formation 
\citep{ns00, vog14, schaye15, nihao16}. It is thus an important test for any theory of galaxy formation and evolution 
to reproduce the slope, scatter and the zero point of the TFr in different 
photometric bands simultaneously. The TFr can also constrain theories about  
the distribution of mass within galaxies, e.g. it was shown by \citet{courteau03} 
that barred and unbarred galaxies follow the same TFr, even though 
barred galaxies could be less dark matter dominated within their optical 
radius \citep{ws01}. 

Over the past decades, the scatter in the observed TFr has been decreased significantly by 
more accurate photometric measures. As first suggested by \citet{aar79}, the TFr can 
be tightened by moving from optical to NIR bands, where the old stars peak in 
luminosity and provide a good proxy for the stellar mass of the galaxies \citep{peletier91}.
The advent of infra-red arrays shifted photometry to the JHK bands and then to 
space-based infrared photometry, e.g. with the {\em Spitzer Space Telescope} 
\citep{wer04}. However, despite the obvious advantages of deep near-infrared 
luminosities, it is still not clear at which NIR wavelengths the smallest scatter in 
the TFr can be achieved. For example, \citet{bern94}, using 23 spirals in the Coma cluster
found that the H-band TFr does not have less scatter than the I-band 
relation. \citet{sor13} claim that the 3.6 $\mu$m TFr has even larger 
scatter than the I-band TFr (\citealt{TC12}, hereafter \citetalias{TC12}).
In fact, accurate infrared photometry led to the point 
where the measurement errors on the total luminosity can no longer explain the observed scatter.  

\begin{figure}
\begin{center}
\includegraphics[scale=0.75]{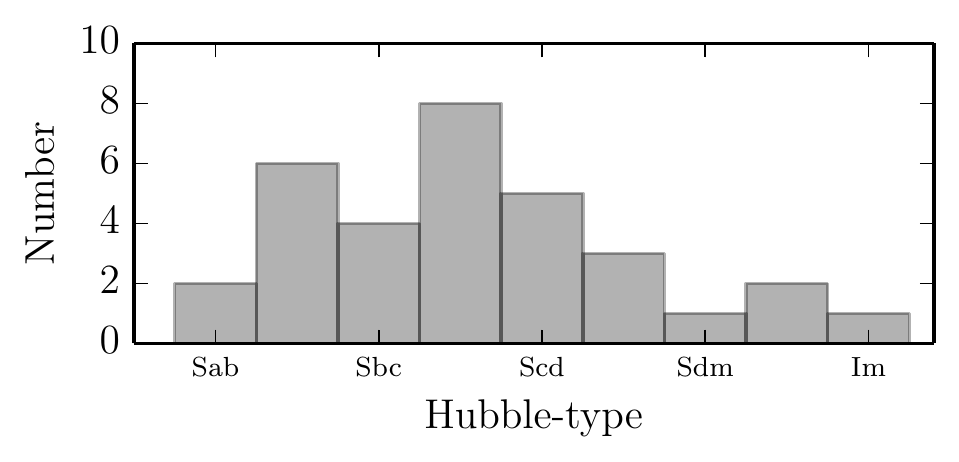}
\caption{
The distribution of morphological types of the galaxies in our sample.
\label{fig_ttype}}
\end{center}
\end{figure}

Yet, sofar, very little attention has been given to improving the measurements of 
the rotation velocity which is, as mentioned before, thought to be strongly related 
dynamically to the dark matter halo. Notably, the width and shape 
of the global H{\sc i} profile are determined by the detailed distribution of the 
H{\sc i} gas in the disk, the shape of a galaxy's rotation curve, and the 
presence of non-circular motions and/or a warp. It is impossible to take into account 
all these important aspects while inferring the rotational velocity of a 
galaxy from the integrated global H{\sc i} profile. This notion has motivated
observational studies which took advantage of optical rotation 
curves, using $H\alpha$ long slit spectroscopy \citep{rubin80,rubin85,pizagno07}. However, the 
rotational velocity at the optical radius does not probe the dark matter halo 
potential properly, since the data do not extend far enough in radius. 
Of course, for an axisymmetric galaxy with a monotonically rising rotation 
curve that reaches a constant flat part in a non-warped outer gas disk, the 
rotational velocity is reasonably well-defined and can be estimated from the 
corrected width of the global H{\sc i} profile. Unfortunately, galaxies often are not
that well-behaved and the column--density distribution and kinematic 
structure of their gas disks may significantly affect the shape and width of the 
global H{\sc i} profiles, introducing errors on the derived rotational velocity 
that can not be corrected for without further information. 

Detailed studies of galaxy rotation 
curves using 21-cm aperture synthesis imaging \citep{bosma81, valbada85, begeman89, broeils94,
v01, swaters02, noord06, deblok14}, 
show that the shape of the rotation curve strongly depends on the morphology and 
surface brightness of the galaxy, introducing deviations from the classical flat 
rotation curve. For instance, it is well--known that late type dwarf galaxies have slowly rising rotation curves. In this case the observed maximal 
rotational velocity ($V_{max}$) will underestimate the velocity of the halo, 
simply because the rotation curve is not reaching the flat part ($V_{flat}$).
The other extreme case are massive early--type spirals, which usually show 
a fast rise of the rotation curve until the maximum velocity ($V_{max}$) is reached,
usually within the optical disk, beyond which the rotation curve significantly declines, reaching 
the flat part with much lower velocity \citep{bosma81,casvangork,v01,noord06}. In this case, the 
mass of the halo, if derived using $V_{max}$, will be overestimated for the 
most massive galaxies, which can cause a curvature in the TFr \citep{neill14}. 
It was shown by \citet{v01} with a study of spiral galaxies in the Ursa Major 
cluster, that the statistical properties of the TFr depend on 
the shape of the rotation curves, and that the observed scatter is reduced 
significantly when extended H{\sc i} rotation curves are available to substitute 
the corrected width of the global H{\sc i} profile with $V_{flat}$ from the rotation curve
as a kinematic measure. 

\begin{table}
\begin{tabular}{ll r@{$\pm$}l r@{$\pm$}l r@{$\pm$}l }
\hline
Name&Hubble type&\multicolumn{2}{l}{P.A.}&\multicolumn{2}{l}{Incl.}&\multicolumn{2}{l}{Distance}\\
&  &\multicolumn{2}{l}{deg.}&\multicolumn{2}{l}{deg.}&\multicolumn{2}{l}{Mpc}\\
\hline
NGC 0055 &SB(s)m        &110&3	 &78&7 &1.98&0.05\\
NGC 0224 &SA(s)b       &37&1	 &78&1 &0.76&0.02\\
NGC 0247 &SAB(s)d       &169&3	 &77&2 &3.51&0.09\\
NGC 0253 &SAB(s)c       &230&2	 &77&1 &3.56&0.13\\
NGC 0300 &SA(s)d       &290&3	 &46&6 &1.97&0.05\\
NGC 0925 &SAB(s)d       &283&2	 &61&5 &8.91&0.28\\
NGC 1365 &SB(s)b       &218&2	 &39&8 &17.7&0.81\\
NGC 2366 &IB(s)m       &42&6	 &68&5 &3.34&0.09\\
NGC 2403 &SAB(s)cd      &124&1	 &61&3 &3.17&0.08\\
NGC 2541 &SA(s)cd      &170&3	 &64&4 &11.5&0.47\\
NGC 2841 &SA(r)b      &150&3	 &70&2 &14.5&0.47\\
NGC 2976 &SAc pec      &323&1   &61&5 &3.63&0.13\\
NGC 3031 &SA(s)ab      &330&4	 &59&5 &3.61&0.09\\
NGC 3109 &SB(s)m        &92&3	 &80&4 &1.37&0.03\\
NGC 3198 &SB(rs)c      &215&5	 &70&1 &13.3&0.55\\
IC 2574  &SAB(s)m      &55&5	&65&10 &3.89&0.14\\
NGC 3319 &SB(rs)cd     &33&2    &57&4 &13.0&0.53\\
NGC 3351 &SB(r)b       &192&1	 &47&5 &10.4&0.28\\
NGC 3370 &SA(s)c       &327&3	 &55&5 &26.1&0.72\\
NGC 3621 &SA(s)d       &344&4	 &65&7 &6.72&0.18\\
NGC 3627 &SAB(s)b       &172&1	 &58&5 &9.03&0.29\\
NGC 4244 &SA(s)cd      &222&1	 &88&3 &4.61&0.19\\
NGC 4258 &SAB(s)bc      &331&1	 &72&3 &7.31&0.16\\
NGC 4414 &SA(rs)c?     &160&2	 &52&4 &17.8&0.74\\
NGC 4535 &SAB(s)c       &180&1	 &41&5 &16.1&0.66\\
NGC 4536 &SAB(rs)bc     &300&3	 &69&4 &14.6&0.60\\
NGC 4605 &SB(s)c pec   &293&2	 &69&5 &5.54&0.25\\
NGC 4639 &SAB(rs)bc     &311&1	 &42&2 &22.0&0.71\\
NGC 4725 &SAB(r)ab pec  &30&3	 &50&5 &12.5&0.46\\
NGC 5584 &SAB(rs)cd     &152&4	 &44&4 &22.4&0.72\\
NGC 7331 &SA(s)b       &169&3	 &75&3 &13.8&0.51\\
NGC 7793 &SA(s)d       &290&2	 &50&3 &3.58&0.11\\
\hline
\end{tabular}
\caption{The Tully-Fisher Calibrator Sample.
Column (1): galaxy name (as shown in \href{https://ned.ipac.caltech.edu/forms/byname.html}{NED}); 
Column (2): Hubble type (as shown in \href{https://ned.ipac.caltech.edu/forms/byname.html}{NED});
Column (3): kinematic position angle (Paper I);
Column (4): kinematic inclination (Paper I);
Column (5): Distance in Mpc provided by The Extragalactic Distance Database 
 \href{http://edd.ifa.hawaii.edu/}{(EDD),} \citealp{EDD}.
}
\label{tbl_samp}
\end{table}

It is very important to realise that the literature contains many 
observational results on the TFr which are often inconsistent with each other. 
This is largely due to different corrections applied to the observables, e.g. for extinction or 
inclination, due to different photometric systems, due to different observing 
techniques or due to different samples. This makes it very complicated to compare the various studies in a simple manner. 
In this paper we establish TFrs based on a homogeneous
analysis of imaging data obtained in 12 photometric bands from UV to IR (the detailed photometric analysis will be presented in a companion paper), 
while taking advantage of spatially resolved H{\sc i} kinematics as reported in \citet{PaperI}, 
hereafter \citetalias{PaperI}.
 We study the statistical properties of the TFr 
to investigate the link between the host dark matter DM halo and the various 
stellar populations of galaxies, which peak in different bands. Such a homogeneous analysis 
allows us to obtain a better understanding of the physical phenomenon of the TFr, 
especially as a tool to study the internal structure of galaxies. 

This paper is organised as follows: Section 2 describes the sample of calibrator 
galaxies. Section 3 describes the collected photometric data. Section 
4 describes the H{\sc i} data.
Section 5 summarises the corrections which were applied to the observables.
Section 6 discusses the statistical properties of the constructed TFrs. Section 7
presents the summary and concluding remarks.

\section{The Sample}
In our study we are interested in the slope and intrinsic tightness of the TFr. We are not 
aiming to maximise the number of galaxies in the sample, but rather to increase the 
quality of the kinematic measures for a representative sample of galaxies with independent distance measurements.
 Thus, we analysed aperture synthesis imaging H{\sc i} data to derive high--quality rotation curves \citepalias{PaperI}. 
The independent distances to our galaxies were measured 
from the Cepheid period--luminosity relation \citep{freedman01} or/and from the tip of the red giant branch \citep{rizzi07} 
and are provided by \href{http://edd.ifa.hawaii.edu/}{The Extragalactic Distance Database (EDD)} \citep{EDD}.
Independently measured distances reduce the error in the absolute magnitude of a galaxy and 
therefore reduce the impact of distance uncertainties on the observed scatter of the TFr. For example, 
in our case, distance uncertainties contribute only $\sigma_{dist}=0.07$ mag to the total observed scatter of the TFr,
which is much lower in comparison with $\sigma_{dist}=0.41$ mag if the Hubble--flow distances are adopted.
We adopt  a sample of 32 large, relatively nearby galaxies from the zero point (ZP) calibrator sample described 
in \citetalias{TC12}. Their selection criteria for galaxies included in the 
sample completely satisfy our requirements: 1) morphological types Sab and later 
(Figure \ref{fig_ttype}),  2) inclinations no less than 45\degree, 3) H{\sc i} 
profiles with adequate S/N, 4) global H{\sc i} profiles without evidence of 
distortion or blending. Their selection criteria give us confidence that the adopted galaxies are kinematically 
well-behaved with regularly rotating, extended H{\sc i} disks. 
In \citetalias{PaperI} it was found that this confidence was largely justified, 
but that the corrected width of the global H{\sc i} profile is not always an accurate representation of $V_{flat}$. 
 Global parameters of the sample galaxies are summarised in Table \ref{tbl_samp}.

\section{Photometric data}
To study the wavelength dependence of the statistical properties of the
TFr requires not only a representative sample, but a systematic, homogeneous 
approach in deriving the main photometric properties of galaxies.
We use 12 bands per galaxy ({\it FUV, NUV, u, g, r, i, z, J, H, K$_{s}$, 3.6, 
4.5 $\mu$m}) for 21 galaxies, and 7 bands for the remaining 11 galaxies in our sample
due to the absence of SDSS imaging data. This broad wavelength coverage allows 
us to measure the relative luminosity of old and young stars within a galaxy and to 
perform SED fits in a forthcoming paper to estimate their stellar masses. 

 The $FUV$ and $NUV$  images were collected from the various  
Galaxy Evolution Explorer (GALEX, \citealp{galex}) space telescope data archives.
 Since young, stars which peak in UV, have a very low contribution to the total mass of a 
 galaxy, a very large scatter in the UV--based TFrs might be expected. 
 Nevertheless, we consider these bands in our study as well.
 
To obtain optical photometry, we use the SDSS Data Release 9 
(DR9, \citealp{sdss}), but note that SDSS photometry is only available for 21 galaxies 
in our sample. Therefore we consider TFrs in the SDSS bands for a smaller number of galaxies. 
However, we point out that the SDSS--subsample still spans a wide luminosity
range.  

We collected a wide range of NIR images:  $J$, $H$, $K_{s}$ bands from the Two-Micron All Sky 
Survey (2MASS, \citealp{2mass}) and 3.6 $\mu$m and 4.5 $\mu$m from the Spitzer Survey for Stellar 
Structure in Galaxies ($S^{4}G$, \citealp{sheth10}). All data were gathered from 
the \href{http://hachi.ipac.caltech.edu:8080/montage}{IRSA archive}. 

\begin{figure}
\begin{center}
\includegraphics[scale=0.5]{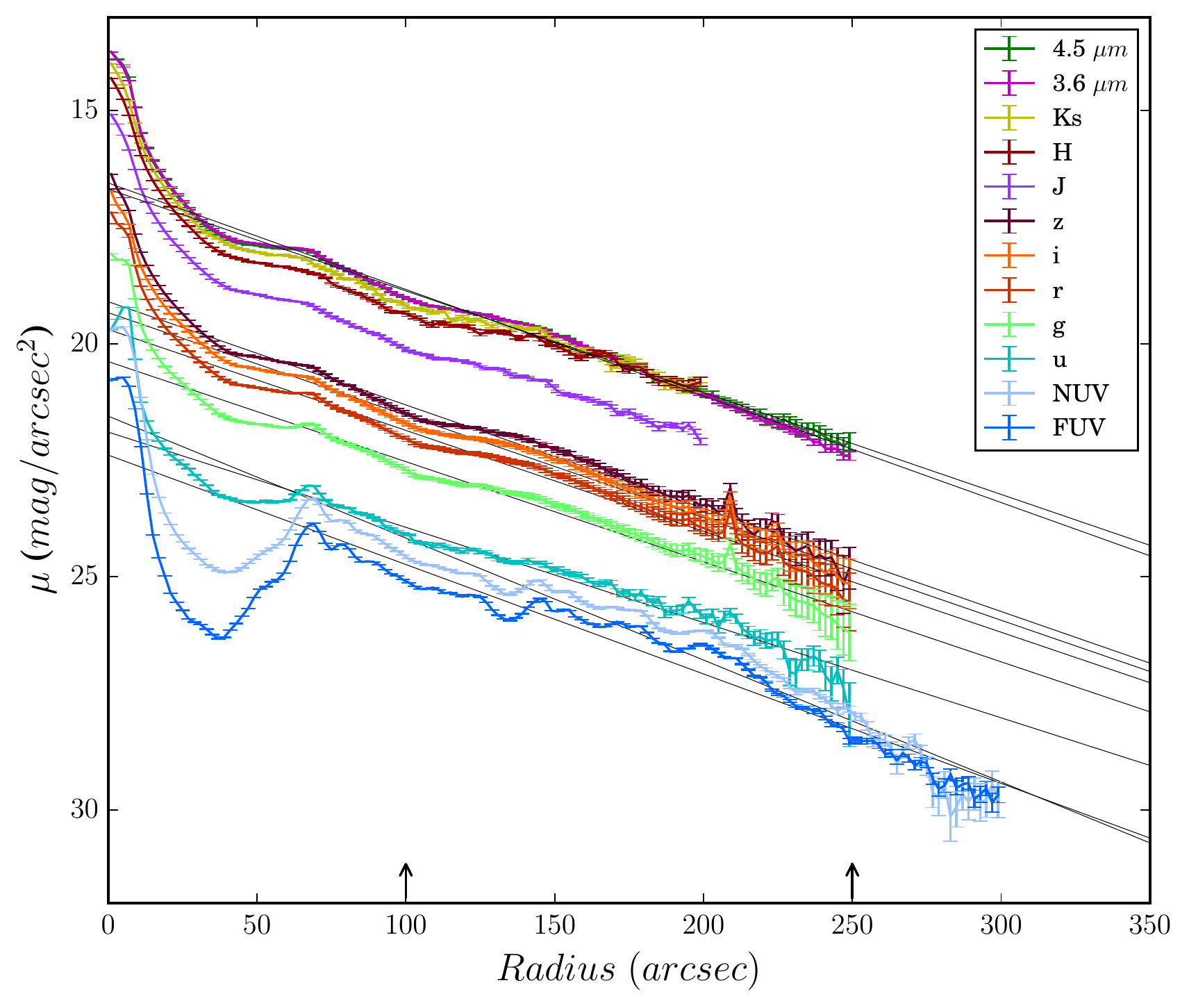}
\caption{
Surface brightness profiles of NGC 3351 for 12 photometric bands. The region 
within which the exponential disk fit was done is indicated with 
arrows. Black lines show the exponential disk fit to the profile. Profiles are terminated at their $R_{lim}$.
\label{fig_sbp}}
\end{center}
\end{figure}

 \subsection{Total Magnitudes}
All data were homogeneously analysed and the total magnitudes were derived for every galaxy in each band.

First, the aperture magnitudes ($m_{ap}$) were calculated by integrating
the surface brightness profile (SBP) in each band within a fixed radius ($R_{lim}$). 
 $R_{lim}$ was chosen after a visual inspection of every profile, as 
the largest radius at which the surface brightness ($\mu_{lim}$) is still reliable.
Given the varying quality of the data, $R_{lim}$ may differ for various bands.
In Figure \ref{fig_sbp} the SBPs are shown for NGC 3351 in all bands within their $R_{lim}$.

Then, linear fits were made to the outer part of each SBP (except for the 2MASS data\footnote[1]{
2MASS survey images are too shallow and the SBPs do not extend enough to apply
the method of recovering the total light as described above. 
Therefore, for $J, H, K$ and 3.6 $\mu$m bands we constructed the growth curves and measured the colours
$J-[3.6]$, $H-[3.6]$ and $K-[3.6]$ at the last reliable measured point of the growth curves of the 2MASS 
bands within the same radius. We fixed these colours and calculated 2MASS total magnitudes as
$M_{tot}^{JHK}=M_{tot}^{[3.6]}+(JHK_{R_{lmp}}-[3.6]_{R_{lmp}})$, where $R_{lmp}$ is the radius 
at which the last measure point of the 2MASS band was measured.}), which 
characterises the exponential drop of surface brightness of a galaxy due to the disk component.
The radial range within which the fit was made was identified through visual inspection 
and is shown in Figure \ref{fig_sbp} with vertical arrows. This procedure is in essence the ``mark the disk" procedure described 
by \citet{dejong96}.
We assume that beyond $R_{lim}$ the SBP continues to drop exponentially without any truncations and/or breaks. 
Under this assumption \citet{tully96} showed that the extended magnitude does not depend on
 the scale length of the disk or on the ellipticity of the galaxy, but only on 
the number $\Delta n$ of disk scale lengths within $R_{lim}$. Hence it can be calculated as :
\begin{equation}
\Delta m_{ext} = 2.5\log[1-(1+\Delta n)e^{-\Delta n}],
\end{equation}
with
\begin{equation}
\Delta n = (\mu_{lim} - \mu_{0})/1.086.
\end{equation}
Then, the total magnitudes follow from $m_{tot} = m_{ap} + m_{ext}$.

Note that the detailed photometric analysis and final data products will be presented in a companion paper. 



\begin{figure}
\begin{center}
\includegraphics[scale=0.55]{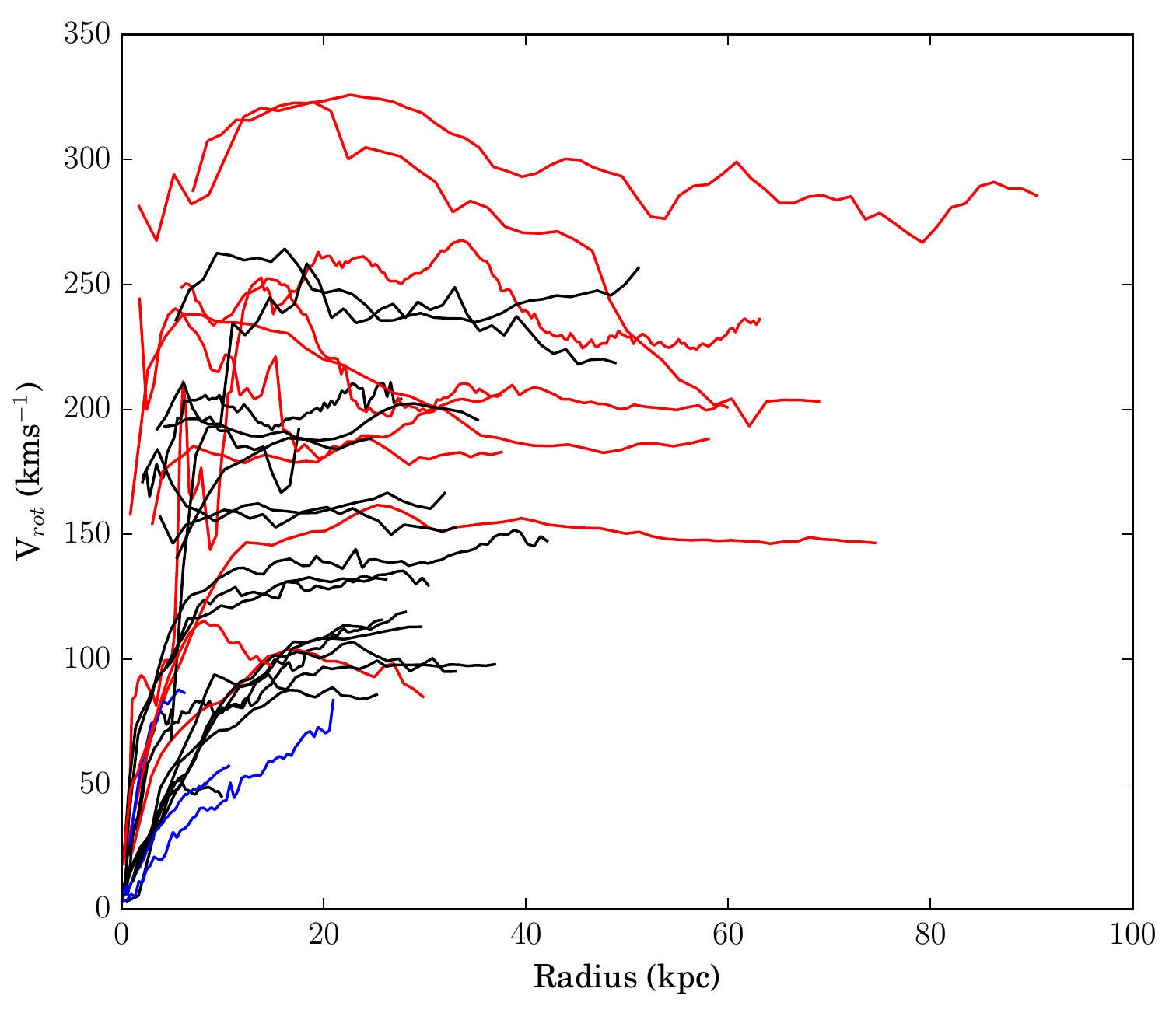}
\caption{Compilation of extended H{\sc i} rotation curves of our sample galaxies plotted on the same
 linear scale. Blue curves belong to galaxies with Rrc ($V_{max} < V_{flat}$) 
and red curves are declining rotation curves ($V_{max} > V_{flat}$).
\label{fig_allcurves}}
\end{center}
\end{figure}
 
 \begin{figure}
\begin{center}
\includegraphics[scale=0.75]{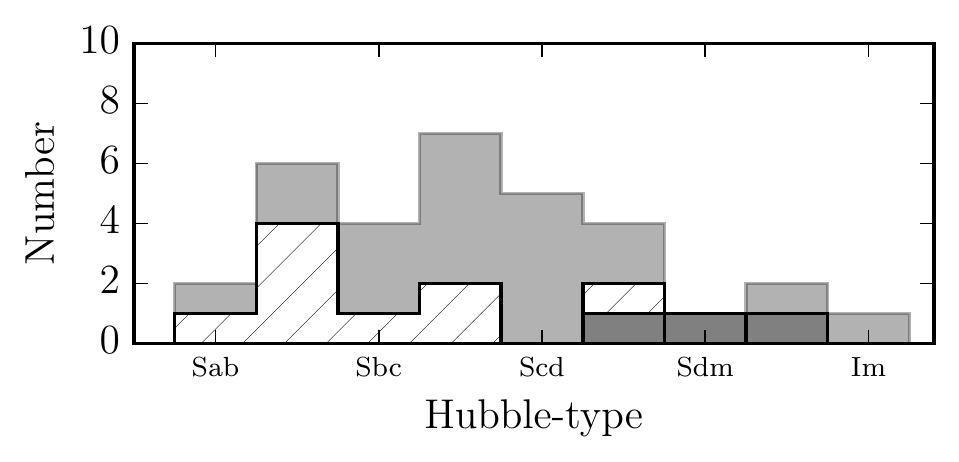}
\caption{
Rotation curves morphology distribution within the sample. The light hatched 
region shows galaxies with declining rotation curves. The dark shaded area 
corresponds to galaxies with rising rotation curves.
\label{fig_rtype}}
\end{center}
\end{figure}

\begin{figure}
\begin{center}
\includegraphics[scale=0.60]{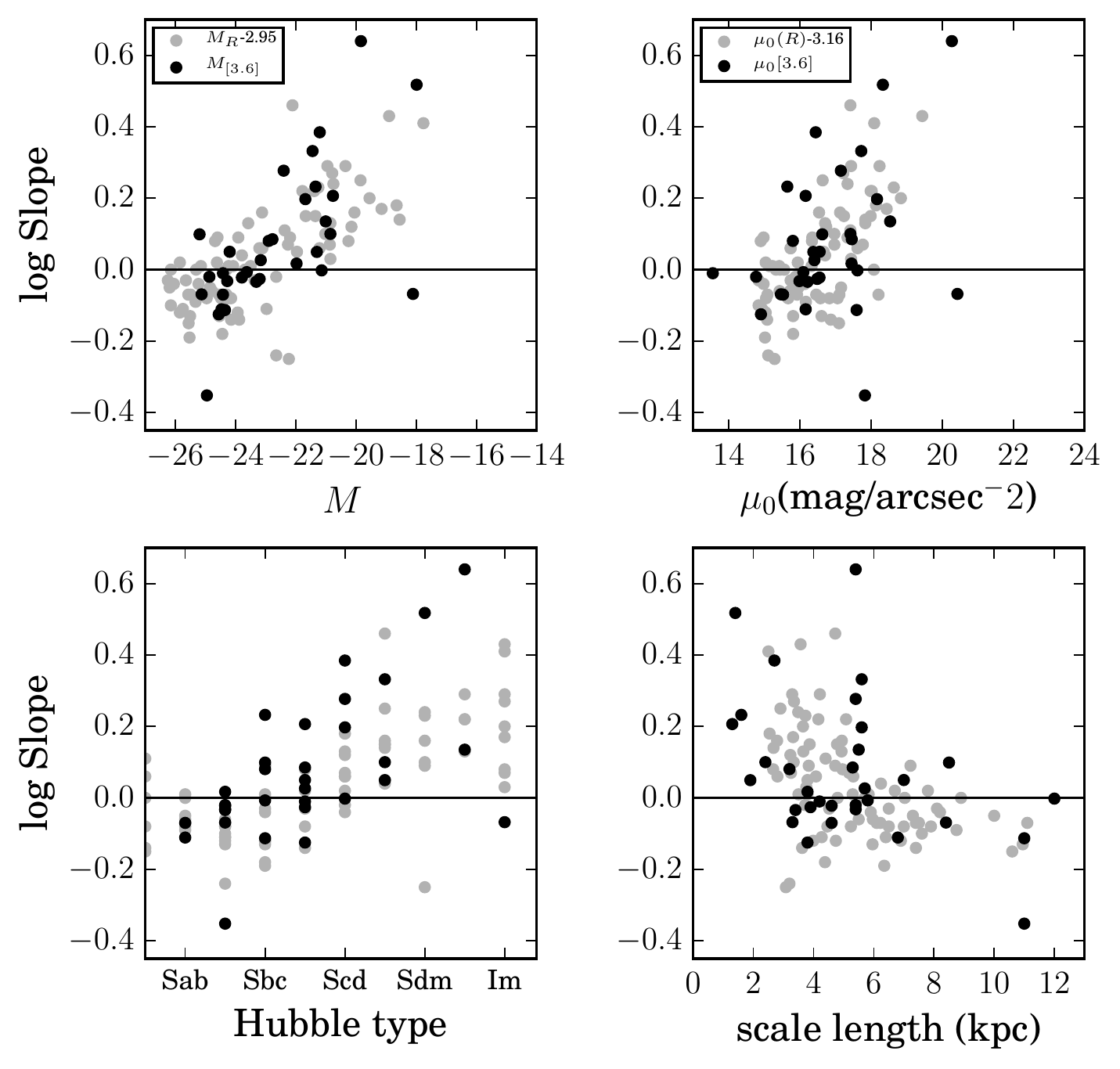}
\caption{ Slopes of the outer part of rotation curves and their correlation 
with galaxy parameters. Our sample is shown with black dots. Absolute magnitude and disk central surface brightness are measured in the
3.6 $\mu$m band. A compilation of various observational samples \citep{casvangork,versanc01, spek06, noord07,swaters09} is shown with grey symbols. For the reference samples, the absolute
 magnitude and disk central surface brightness are measured in the $R$ band and then matched to our sample using the colour term ($R-[3.6]=2.95$
 and $\mu_{0}(R)-\mu_{0}[3.6]=3.16$). 
\label{fig_slopes}}
\end{center}
\end{figure}

\section {H{\sc i} data} 
We are interested in studying the statistical properties of the TFr at various wavelengths
based on rotational velocities derived from global H{\sc i} profiles and high--quality rotation curves. 
The ideal data for this work are H{\sc i} radio aperture synthesis imaging data which provide the global H{\sc i} profiles,
as well as the spatially resolved rotation curves.

We collected the H{\sc i} radio synthesis--imaging data for 29 galaxies from the literature.
Most of these galaxies were observed previously as part of larger H{\sc i} surveys (THINGS, WHISP, HALOGAS, etc). 
We observed the remaining three galaxies with the Giant Radio Metrewave Telescope (GMRT) in March 2014. 
All data cubes were analysed homogeneously and the following data products were delivered  for all 32 galaxies in our sample:
global H{\sc i} profiles, integrated H{\sc i} column--density maps, H{\sc i} surface--density profiles and
high--quality rotation curves derived from highly--resolved, two--dimensional velocity fields. 

These data products, along with a detailed description of the observations, data reduction and analyses, 
are presented in \citetalias{PaperI}.
Here, we summarize the relevant kinematic information, obtained from these H{\sc i} data.

\begin{table}
\begin{tabular}{lr@{$\pm$}llr@{$\pm$}llr}
\hline
Name&\multicolumn{2}{l}{$V_{sys}$}  &$W_{50}^{i}$    &\multicolumn{2}{l}{$V_{max}$}   & $V_{flat}   $& Slope  \\
&\multicolumn{2}{l}{$km s^{-1}$}&km s$^{-1}$&\multicolumn{2}{l}{$km s^{-1}$}&$km s^{-1}$&\\
\hline
NGC 0055 &130&5  &185$\pm$4   &85 &1	&85$\pm$2  & 0.135  \\
NGC 0224 &-300&3 &517$\pm$5   &261&2    &230$\pm$7 &-0.02  \\
NGC 0247 &160&10 &200$\pm$3   &110&5    &110$\pm$5&0.332 \\
NGC 0253 &240&5	 &410$\pm$3   &200&4    &200$\pm$4&	-0.01   \\
NGC 0300 &135&10 &160$\pm$5   &103&3	&85$\pm$7  & 0.100  \\
NGC 0925 &550&5	 &200$\pm$8   &115&4	&115$\pm$4 & 0.277 \\
NGC 1365 &1640&3 &380$\pm$10  &322&6    &215$\pm$4 &-0.352 \\
NGC 2366 &107&10 &100$\pm$10  &45&5      &45$\pm$5 &-0.068   \\
NGC 2403 &135&1	 &225$\pm$1   &128&1    &128$\pm$1 & 0.017 \\
NGC 2541 &560&5	 &200$\pm$6   &100&4    &100$\pm$4 & 0.384 \\
NGC 2841 &640&20 &590$\pm$3   &325&2    &290$\pm$6 &-0.069  \\
NGC 2976 &5&5  &130$\pm$7 	  &78&4      &78$\pm$4 & 0.206  \\
NGC 3031 &-40&10 &415$\pm$6   &249&3    &215$\pm$9 &-0.070   \\
NGC 3109 &404&5	 &110$\pm$1   &57&2      &  --	   & 0.517  \\
NGC 3198 &660&10 &315$\pm$4   &161&2 	&154$\pm$4 & 0.026   \\
IC 2574  &51&3 &105$\pm$2   &75&5       &  --	   & 0.639     \\
NGC 3319 &730&4	 &195$\pm$6   &112&10 &112$\pm$10  & 0.197  \\
NGC 3351 &780&5  &260$\pm$6   &190&5 &176$\pm$8	   &-0.022   \\
NGC 3370 &1280&15&275$\pm$4   &152&4 &152$\pm$4	   &-0.026    \\
NGC 3621 &730&13 &275$\pm$7   &145&   &145$\pm$5   & 0.080    \\
NGC 3627 &715&10 &340$\pm$7   &183&7	 &183$\pm$7&-0.032    \\
NGC 4244 &245&3	 &195$\pm$6   &110&6	 &110$\pm$6&-0.002   \\
NGC 4258 &445&15 &420$\pm$6   &242&5 	&200$\pm$5 &-0.113   \\
NGC 4414 &715&7	 &375$\pm$6   &237&10	&185$\pm$10&-0.125	   \\
NGC 4535 &1965&5 &270$\pm$6   &195&4	 &195$\pm$4& 0.050    \\
NGC 4536 &1800&6 &320$\pm$6   &161&10	&161$\pm$10&-0.007	    \\
NGC 4605 &160&15 &150$\pm$15  &87&4       &  --    & 0.232       \\
NGC 4639 &978&20 &275$\pm$6   &188&1	 &188$\pm$1&-0.034       \\
NGC 4725 &1220&14&400$\pm$4   &215&   &215$\pm$5   &-0.111    \\
NGC 5584 &1640&6 &190$\pm$10  &132&2  &132$\pm$2   & 0.085     \\
NGC 7331 &815&5	 &500$\pm$10  &275&5  	&275$\pm$5&	 0.098        \\
NGC 7793 &228&70 &174$\pm$10  &118&8  &95$\pm$8    & 0.049     \\
\hline
\end{tabular}
\caption{Results from the H{\sc i} kinematics analysis. 
Column (1): galaxy name;
 Column (2): systemic velocity;
Column (3): width of the global H{\sc i} profile at 50\% level;
Column (4): maximal rotational velocity;
Column (5): rotational velocity of the flat part of rotation curve.
Column (6): slope of the outer part of the rotation curve.
}
\label{tbl_rot}
\end{table}

\begin{figure*}
\includegraphics[scale=0.55]{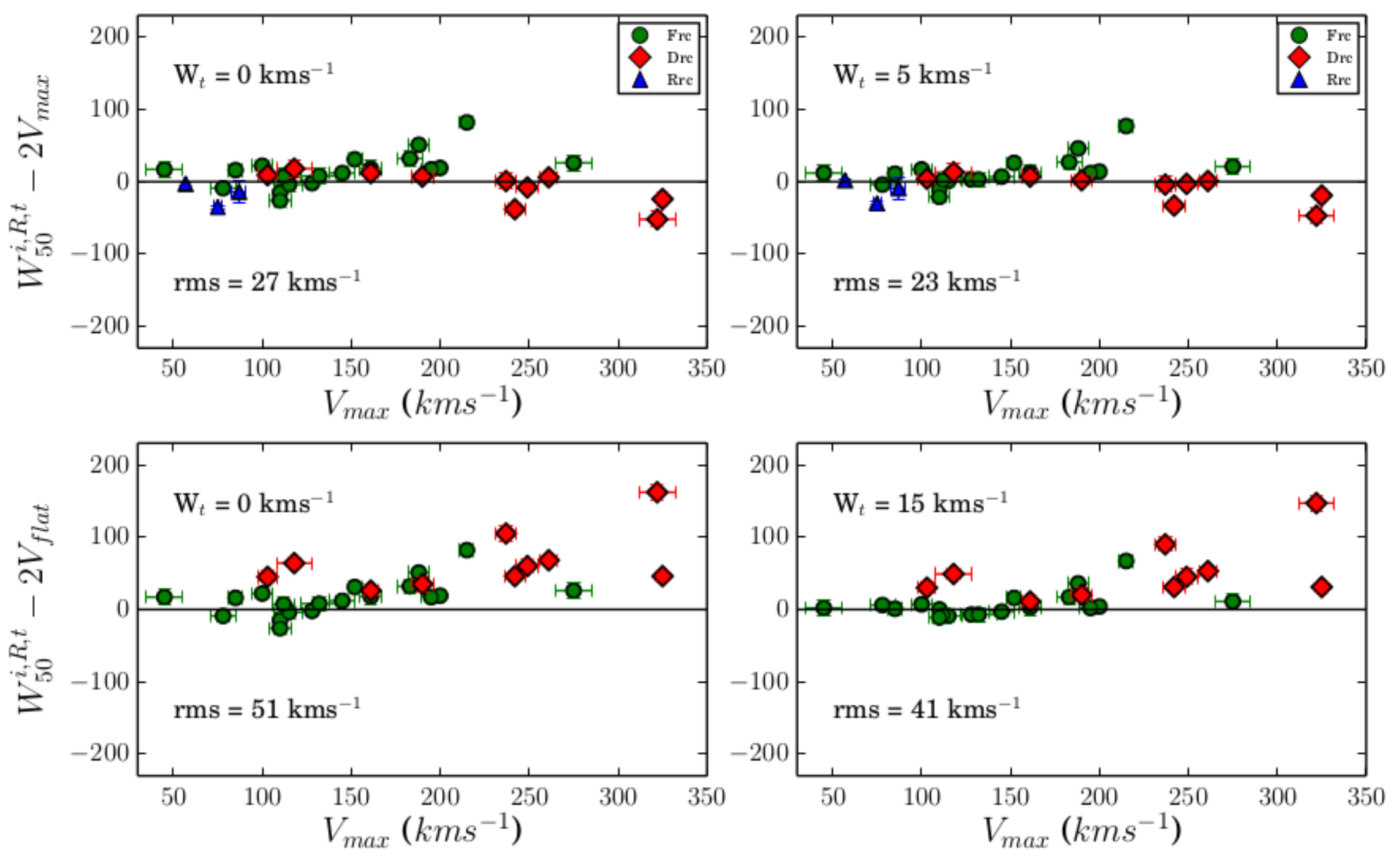}
\caption{Comparison of the global profile widths at the 50\% level, corrected 
for inclination, instrumental broadening and random motions $W_{50}^{i,R,t}$ 
with $2V_{max}$ (upper panels) and $2V_{flat}$ (bottom panels). Blue symbols 
indicate galaxies with Rrc ($V_{max} < V_{flat}$) and red symbols indicate 
galaxies with declining rotation curves ($V_{max} > V_{flat}$). Green symbols 
show flat rotation curves ($V_{max} = V_{flat}$).
\label{fig_vres}}
\end{figure*}

\subsection {Rotational velocities}
There are several methods to measure the rotational velocities of spiral galaxies using H{\sc i} data. 
First,  from the width of the H{\sc i} 21cm line profile, where 
the corrected width of the profile relates to the rotational velocity as 
$W^{i}=2V_{rot}$. Second, from the spatially resolved H{\sc i} velocity fields which allow the derivation of the high--quality 
rotation curves.  
While the former is much faster to obtain observationally with single--dish telescopes and can therefore be 
used for a large number of galaxies, the latter provides valuable extra information.

Amongst others, \citet{v01} showed that rotation curves of spiral galaxies have different 
shapes, which mostly depend on the morphological type and luminosity of a galaxy.
The advantage of our sample is that it covers all types of rotation curves: 
rising (Rrc) for dwarf galaxies ($V_{max} < V_{flat}$), classical flat (Frc) for 
the intermediate types ($V_{max} = V_{flat}$), and declining (Drc) for the 
early--type spirals ($V_{max} > V_{flat}$). The ``family'' of rotation 
curves of our sample is shown in Figure \ref{fig_allcurves}. 
Moreover, Figure \ref{fig_rtype} demonstrates that declining rotation curves tend to belong mainly to massive early--type 
spirals, while rising rotation curves are common for late type galaxies \citep{casvangork, v01,noord07, oh08, swaters09, deblok14}. 

To quantitatively describe the shape of a rotation curve, we 
measure the slope of the outer part of a rotation curve between the radius at 2.2 disk 
scale lengths, measured from the 3.6.$\mu$m surface brightness profile (SBP), and the outermost point:
\begin{equation} 
S_{2.2h,lmp} = \frac{Log(V_{2.2h}/V_{lmp})}{Log(R_{2.2h}/R_{lmp})}
\end{equation}
where $V_{2.2h}$ and $V_{lmp}$ are the rotational velocities at the radius 
equal to $2.2h$ $(R_{2.2h})$ and at the radius of the last measured 
point ($R_{lmp}$). Thus, a slope equal to zero belongs to a flat rotation curve, 
a positive slope to a rising rotation curve and a negative slope to a declining rotation curve.
Figure \ref{fig_slopes} demonstrates $S_{2.2h,lmp}$ as 
a function of global galactic properties. There are prominent 
correlations with Hubble type, as was suggested earlier (Figure \ref{fig_rtype}), and with the absolute 
magnitude. In grey we show a compilation of various samples from 
previous studies \citep{casvangork,versanc01, spek06, noord07,swaters09},
to point out that our sample is not in any way peculiar and follows the same trends
found in previous studies. 

The shape of the global H{\sc i} profile may hint at
the shape of the rotation curve in several cases. First, a 
boxy or Gaussian shape profile is an indication for a rising rotation curve for
which the velocity of the dark matter halo is underestimated from the profile width.
In these profiles, the classical double peak is missing because the constant, flat part of the rotation curve 
is not sampled by an extended H{\sc i} disk. In its turn, 
the classical double peak profile gives an indication that the rotation curve of 
a galaxy will reach its flat part. 
Figure \ref{fig_vres} demonstrates the difference in the velocity obtained using 
$W_{50}^{i}$ (corrected for inclination), compared to $V_{max}$ (upper panel) and $V_{flat}$ 
(bottom panel) as derived from the rotation curves. It is clear that the main outliers are 
the galaxies with either rising (blue) or declining (red) rotation curves. 
Thus, the rotational velocity measured from $W_{50}^{i}$ will be underestimated in comparison with 
 $V_{max}$, and overestimated in comparison with 
$V_{flat}$. Therefore, one should take into account that the rotational velocity derived
from the width of the global profile may differ from the velocity measured from the spatially resolved rotation curve.

\section {Corrections to observables}
\subsection {Photometry}
As a photometric measure for the TFr we use the corrected absolute total 
magnitudes $M_{T}^{b,i}(\lambda)$ :
\begin{equation} 
M_{T}^{b,i}(\lambda) = m_{T}(\lambda) - A^{i}_{\lambda} - A^{b}_{\lambda} - DM,
\end{equation}
where $m_{T}(\lambda)$ is the apparent total magnitude (Section 3.1), $A^{\lambda}_{i}$ is 
the internal extinction correction, $A^{\lambda}_{b}$ is the Galactic extinction 
correction and DM is the distance modulus, based on the distance given in Table \ref{tbl_samp}. 
It is important to note, that we not only measure the total magnitudes for all galaxies in each band 
in the same manner, we also use the same methods to apply corrections due to the Galactic and internal extinction, 
therefore we can perform a fair comparisons between the TFrs in various bands. The detailed 
analysis of the photometrical corrections can be found in a forthcoming companion paper. 

\subsection { H{\sc i} kinematics}
In this section we summarize the main corrections that were applied to the kinematic measures. 

 The global H{\sc i} linewidths were corrected for: 
\begin{enumerate}
\item Instrumental broadening, which depends on the instrumental velocity resolution and on the steepness of the wings of the velocity profile,
following \citet{versanc01}.
\item Turbulent motions, which depend on the level at which the width of the profile was measured (20\% or 50\%
of the peak flux, \citealp{versanc01}). 
Figure \ref{fig_vres} demonstrates the difference between the corrected (right panels) and non-corrected (left panels)
widths of the integrated H{\sc i} profile and the velocity derived from the rotation curve.
\item Inclination, according to the formula $W_{50,20}^{i}=W_{50,20}/sin(i_{kin})$, 
where $i_{kin}$ is the kinematic inclination angle derived from the H{\sc i} velocity fields \citepalias{PaperI}, in order to represent the rotational velocity as $W_{50,20}^{i}=2V_{rot}$ (see Section 4.1).
\end{enumerate}

Prior to the rotation curve derivation, the H{\sc i} velocity fields were censored for skewed velocity profiles (high $h3$).
A skewness of the velocity profiles might be present mostly due to beam--smearing and non--circular motions. Thus, censoring for
high values of $h3$ allowed us to derive high--quality rotation curves, representing the actual rotational velocity of a galaxy as a function 
of radius, not affected by 
the effects mentioned above. Further details can be found in \citetalias{PaperI}.

\section {The Tully-Fisher relations}
In this section we present the statistical properties of the 
multi-wavelength TFrs using the different kinematic measures $W_{50}^{i}$, 
$V_{max}$ and $V_{flat}$. We first discuss the fitting 
method. We then discuss the slope and vertical scatter ($\sigma$) of the TFrs and
 introduce the slope independent  tightness ($\sigma_{\perp}$) of the TFrs. 
 We conclude with a search for a 2$^{nd}$ parameter that may correlate with the residuals.

\begin{table*}
\begin{tabular}{lcccccc}
\hline
 Band&\multicolumn{3}{c}{Slope (Mag)}&\multicolumn{3}{c}{Slope (dex)}\\
\hline
 &$W_{50}^{i}$& $V_{max}$ &$V_{flat}$ &$W_{50}^{i}$& $V_{max}$ &$V_{flat}$\\
 \hline
 FUV & -7.12$\pm$0.86 & -7.04$\pm$0.98 &  -7.87$\pm$1.27 & 2.36$\pm$0.38 &2.32$\pm$0.37   &2.59$\pm$0.47 \\    
 NUV & -6.45$\pm$0.76 & -6.36$\pm$0.86 &  -6.91$\pm$0.94 & 1.93$\pm$0.40  &1.91$\pm$0.40  &2.06$\pm$0.41 \\    
 u   & -6.06$\pm$0.64 & -6.30$\pm$0.60 &  -6.95$\pm$0.67 & 1.69$\pm$0.44  &1.75$\pm$0.46  &1.91$\pm$0.52 \\   
 g   & -6.11$\pm$0.33 & -6.43$\pm$0.57 &  -7.12$\pm$0.6  & 1.9 $\pm$0.33  &1.97$\pm$0.4   &2.17$\pm$0.44 \\   
 r   & -6.76$\pm$0.25 & -7.09$\pm$0.51 &  -7.87$\pm$0.56 & 2.26$\pm$0.23  &2.36$\pm$0.3   &2.61$\pm$0.33 \\   
 i   & -7.02$\pm$0.32 & -7.29$\pm$0.49 &  -8.14$\pm$0.57 & 2.12$\pm$0.38  &2.19$\pm$0.44  &2.42$\pm$0.49 \\   
 z   & -7.89$\pm$0.40 & -8.17$\pm$0.52 &  -9.12$\pm$0.61 & 2.82$\pm$0.16  &2.91$\pm$0.22  &3.25$\pm$0.24 \\   
 J   & -8.73$\pm$0.52 & -8.55$\pm$0.39 &  -9.22$\pm$0.4  & 3.23$\pm$0.26  &3.16$\pm$0.20  &3.41$\pm$0.19 \\   
 H   & -8.99$\pm$0.52 & -8.83$\pm$0.42 &  -9.47$\pm$0.38 & 3.43$\pm$0.24  &3.36$\pm$0.18  &3.61$\pm$0.15 \\   
 K$^{s}$   & -9.26$\pm$0.50 & -9.08$\pm$0.41 &  -9.77$\pm$0.41 & 3.51$\pm$0.23  &3.44$\pm$0.18  &3.81$\pm$0.19 \\   
3.6 $\mu$m  &  -9.05$\pm$0.45 & -8.86$\pm$0.37 &  -9.52$\pm$0.32 & 3.61$\pm$0.19  &3.53$\pm$0.15  &3.8 $\pm$0.11 \\   
4.5$\mu$m  & -9.04$\pm$0.46 & -8.81$\pm$0.38 &  -9.51$\pm$0.33 & 3.52$\pm$0.19  &3.45$\pm$0.16  &3.7 $\pm$0.12 \\    
\hline
Band&\multicolumn{3}{c}{Zero Point (Mag)}&\multicolumn{3}{c}{Zero point (log($L(L_{\sun})$))}\\
\hline
 &$W_{50}^{i}$& $V_{max}$ &$V_{flat}$ &$W_{50}^{i}$& $V_{max}$ &$V_{flat}$\\
 \hline
 FUV       &  0.18$\pm$2.17 & -0.09$\pm$2.44 &   1.77$\pm$3.12  & 7.53$\pm$0.95  &7.66$\pm$0.93  &7.05$\pm$1.16 \\    
 NUV       & -1.78$\pm$1.92 & -2.07$\pm$2.15 &  -0.88$\pm$2.32  & 6.36$\pm$1.03  &6.47$\pm$1     &6.11$\pm$1.04 \\    
 u         & -2.45$\pm$1.62 & -1.91$\pm$1.53 &  -0.41$\pm$1.71  & 4.71$\pm$1.13  &4.59$\pm$1.2   &4.21$\pm$1.32 \\   
 g         & -4.27$\pm$0.83 & -3.55$\pm$1.42 &  -1.94$\pm$1.51  & 4.96$\pm$0.86  &4.79$\pm$1.03  &4.33$\pm$1.13 \\   
 r         & -3.24$\pm$0.63 & -2.49$\pm$1.28 &  -0.67$\pm$1.39  & 4.28$\pm$0.59  &4.07$\pm$0.78  &3.49$\pm$0.84 \\   
 i         & -2.94$\pm$0.81 & -2.32$\pm$1.25 &  -0.35$\pm$1.44  & 4.74$\pm$0.99  &4.58$\pm$1.12  &4.04$\pm$1.25 \\   
 z         & -0.87$\pm$1.01 & -0.24$\pm$1.31 &   1.97$\pm$1.52  & 3.07$\pm$0.41  &2.85$\pm$0.56  &2.08$\pm$0.62 \\   
 J         &  0.1 $\pm$1.28 & -0.44$\pm$0.99 &   1   $\pm$0.99  & 1.92$\pm$0.67  &2.13$\pm$0.52  &1.59$\pm$0.49 \\   
 H         &  0.08$\pm$1.29 & -0.42$\pm$1.06 &   0.92$\pm$0.94  & 1.70$\pm$0.61  &1.90$\pm$0.45  &1.38$\pm$0.37 \\   
 K$^{s}$   &  0.52$\pm$1.24 &  0.00$\pm$1.04 &   1.44$\pm$0.038 & 1.58$\pm$0.6   &1.78$\pm$0.46  &1.22$\pm$0.44 \\   
3.6 $\mu$m  &  2.4$\pm$1.1   &  1.91$\pm$1.41 &   3.31$\pm$1.73  & 1.44$\pm$0.49  &1.66$\pm$0.37  &1.1$\pm$0.29 \\   
4.5$\mu$m  &  2.92$\pm$1.13 &  2.36$\pm$0.97 &   3.73$\pm$0.83  & 1.66$\pm$0.50  &1.88$\pm$0.39  &1.33$\pm$0.31 \\  
\hline
\end{tabular}
\caption{The results of the orthogonal fits of the TFrs. 
Upper panel. Column (1): photometric band;
Column (2)-Column(4): slopes of the TFrs based on $W_{50}^{i}$, $V_{max}$ and $V_{flat}$, measured in magnitudes;
Column (5)-Column(7): slope of the TFrs based on $W_{50}^{i}$, $V_{max}$ and $V_{flat}$, measured in dex.
Lower panel. Column (1): photometric band;
Column (2)-Column(4): zero points of the TFrs based on $W_{50}^{i}$, $V_{max}$ and $V_{flat}$, measured in magnitudes;
Column (5)-Column(7): zero points of the TFrs based on $W_{50}^{i}$, $V_{max}$ and $V_{flat}$, measured in dex. }
\label{tbl_slope}
\end{table*}            

\begin{table*}
\begin{tabular}{lllllll}
\hline
 Band&\multicolumn{3}{c}{ $\sigma$ (Mag)}&\multicolumn{3}{c}{ $\sigma$ (dex)}\\
\hline
  &$W_{50}^{i}$& $V_{max}$ &$V_{flat}$ &$W_{50}^{i}$& $V_{max}$ &$V_{flat}$\\
  \hline
 FUV & 0.87$\pm$0.14 & 0.89$\pm$0.15 & 0.97$\pm$0.17 & 0.30$\pm$0.09 & 0.31$\pm$0.09 & 0.33$\pm$0.1   \\    
 NUV & 0.74$\pm$0.13 & 0.76$\pm$0.14 & 0.77$\pm$0.15 & 0.25$\pm$0.09 & 0.26$\pm$0.09 & 0.26$\pm$0.09   \\    
 u   & 0.44$\pm$0.15 & 0.47$\pm$0.14 & 0.44$\pm$0.15 & 0.19$\pm$0.11 & 0.19$\pm$0.12 & 0.19$\pm$0.12   \\   
 g   & 0.27$\pm$0.11 & 0.34$\pm$0.14 & 0.32$\pm$0.14 & 0.14$\pm$0.1 & 0.15$\pm$0.11 & 0.15$\pm$0.11   \\   
 r   & 0.22$\pm$0.09 & 0.31$\pm$0.13 & 0.29$\pm$0.13 & 0.12$\pm$0.08 & 0.14$\pm$0.1  & 0.13$\pm$0.1   \\   
 i   & 0.25$\pm$0.1  & 0.3 $\pm$0.12 & 0.31$\pm$0.13 & 0.11$\pm$0.06 & 0.12$\pm$0.08 & 0.12$\pm$0.08   \\   
 z   & 0.29$\pm$0.11 & 0.32$\pm$0.13 & 0.33$\pm$0.14 & 0.12$\pm$0.07 & 0.13$\pm$0.08 & 0.13$\pm$0.09   \\   
 J   & 0.39$\pm$0.11 & 0.4 $\pm$0.09 & 0.39$\pm$0.09 & 0.15$\pm$0.08 & 0.16$\pm$0.07 & 0.15$\pm$0.06   \\   
 H   & 0.41$\pm$0.11 & 0.44$\pm$0.1  & 0.38$\pm$0.09 & 0.16$\pm$0.07 & 0.17$\pm$0.06 & 0.15$\pm$0.06   \\   
 K   & 0.41$\pm$0.1  & 0.44$\pm$0.1  & 0.40$\pm$0.09 & 0.16$\pm$0.07 & 0.17$\pm$0.06 & 0.16$\pm$0.06   \\   
3.6  & 0.39$\pm$0.1  & 0.41$\pm$0.09 & 0.33$\pm$0.08 & 0.15$\pm$0.06 & 0.16$\pm$0.05 & 0.13$\pm$0.05   \\   
4.5  & 0.40$\pm$0.1  & 0.42$\pm$0.09 & 0.34$\pm$0.08 & 0.16$\pm$0.06 & 0.16$\pm$0.05 & 0.13$\pm$0.05   \\ 
\hline
\end{tabular}
\caption{Vertical scatter of the TFrs  in different photometrical bands measured in magnitudes and in dex. 
Column (1): photometric band;
Column (2)-Column(4): scatters of the TFrs based on $W_{50}^{i}$, $V_{max}$ and $V_{flat}$, measured in magnitudes;
Column (5)-Column(7): tightnesses of the TFrs based on $W_{50}^{i}$, $V_{max}$ and $V_{flat}$, measured in dex.
}
\label{tbl_scatter}
\end{table*}

\begin{table}
\begin{tabular}{llll}
\hline
 Band&\multicolumn{3}{c}{ $\sigma_{\perp}$ (dex)}\\
\hline
&$W_{50}^{i}$& $V_{max}$ &$V_{flat}$ \\
  \hline
 FUV &0.12$\pm$0.045 & 0.124$\pm$0.022 & 0.122$\pm$0.022  \\
 NUV &0.12$\pm$0.048 & 0.123$\pm$0.024 & 0.117$\pm$0.024  \\ 
 u   &0.099$\pm$0.062 & 0.099$\pm$0.031 & 0.091$\pm$0.031  \\ 
 g   &0.068$\pm$0.044 & 0.071$\pm$0.022 & 0.066$\pm$0.022  \\ 
 r   &0.05$\pm$0.03 & 0.055$\pm$0.015 & 0.05$\pm$0.015  \\ 
 i   &0.042$\pm$0.018 & 0.044$\pm$0.009 & 0.041$\pm$0.009  \\ 
 z   &0.043$\pm$0.02 & 0.045$\pm$0.01 & 0.042$\pm$0.01  \\ 
 J   &0.047$\pm$0.016 & 0.049$\pm$0.013 & 0.045$\pm$0.013  \\ 
 H   &0.047$\pm$0.015 & 0.050$\pm$0.012 & 0.042$\pm$0.012  \\ 
 K   &0.046$\pm$0.022 & 0.049$\pm$0.011 & 0.042$\pm$0.011  \\ 
3.6  &0.043$\pm$0.019 & 0.046$\pm$0.01 & 0.036$\pm$0.01  \\ 
4.5  &0.044$\pm$0.02 & 0.047$\pm$0.01 & 0.036$\pm$0.01  \\ 
 
\hline
\end{tabular}
\caption{Tightness of the TFrs in different photometric bands measured in dex. 
Column (1): photometric band;
Column (2)-Column(4): tightness of the TFrs based on $W_{50}^{i}$, $V_{max}$ and $V_{flat}$, measured in dex;
}
\label{tbl_tght}
\end{table}  

\begin{figure}
\begin{center}
\includegraphics[scale=0.75]{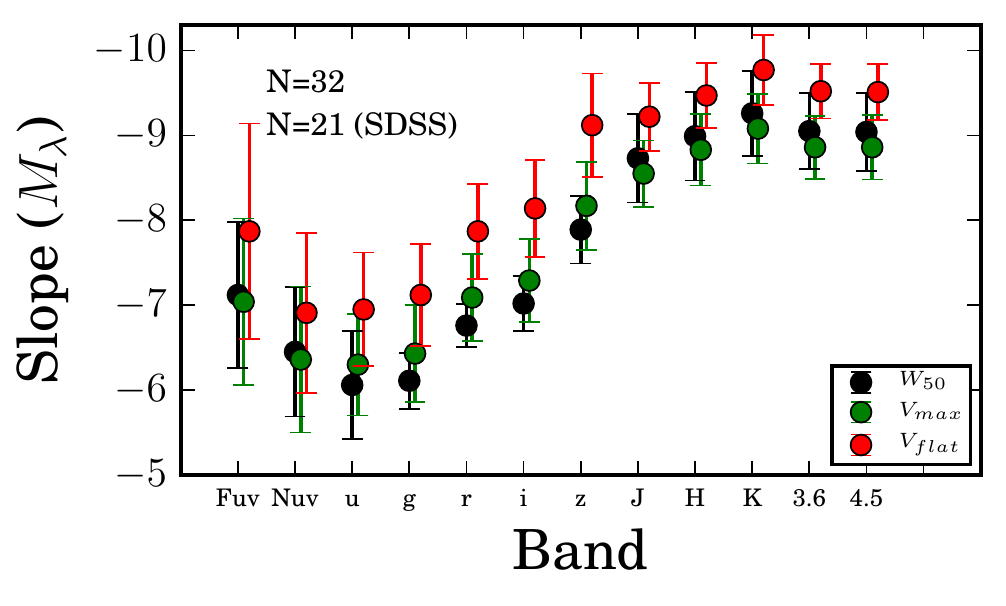}
\caption{
Slope of the TFr as a function of wavelength, calculated using 
different rotation measures. With black points indicated slopes measured for 
the TFr based on $W_{50}^{i}$, with green based on $V_{max}$ and with red based on $V_{flat}$. 
Independently of band, the Tfr based on $V_{flat}$  demonstrates the 
steepest slope.
\label{fig_slope}}
\end{center}
\end{figure}

\begin{figure}
\begin{center}
\includegraphics[scale=0.75]{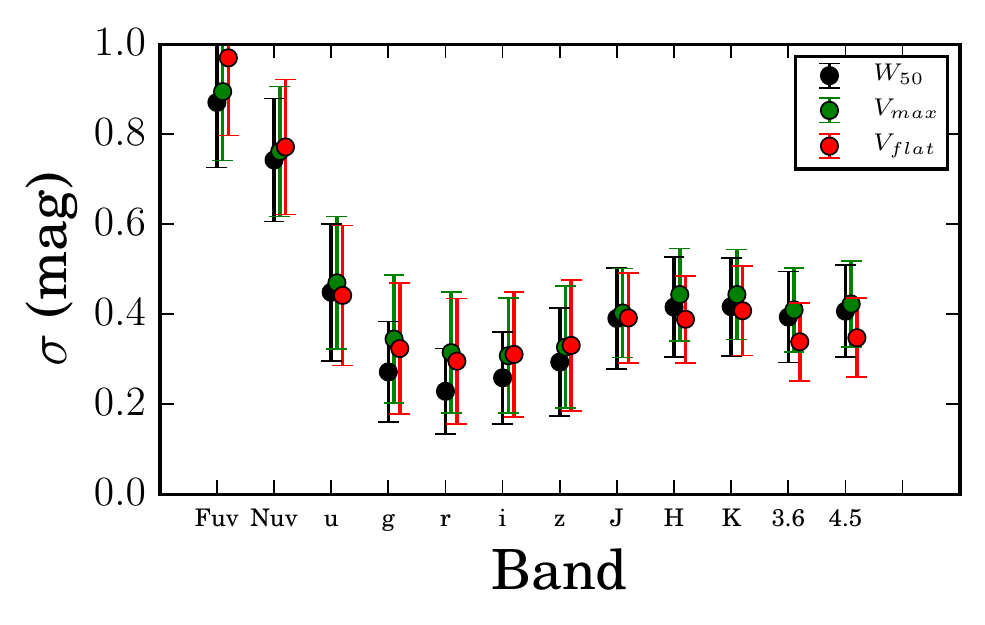}
\caption{
The vertical scatter of the TFr as a function of wavelength, calculated using 
different rotation measures. With black points indicating the scatter measured for 
the TFr based on $W_{50}^{i}$, with green based on $V_{max}$ and with red based on $V_{flat}$. 
\label{fig_scat}}
\end{center}
\end{figure}

\begin{figure*}
\begin{center}
\includegraphics[scale=0.65]{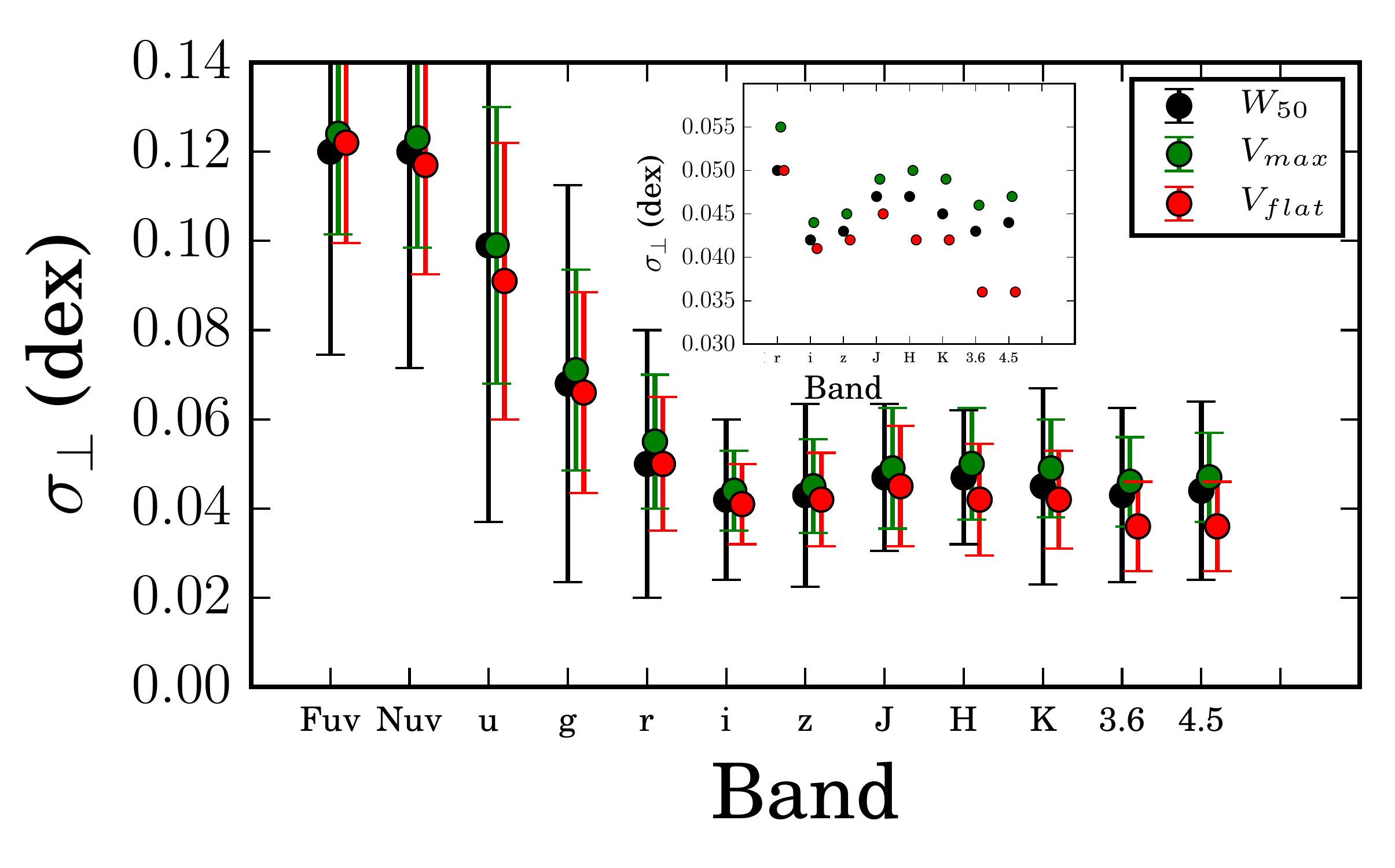}
\caption{
Orthogonal tightness of the TFr as a function of wavelength, calculated using 
different rotation measures. With black points indicated the scatter measured for 
the TFr based on $W_{50}^{i}$, with green based on $V_{max}$ and with red based on $V_{flat}$. 
Independently of wavelength the relation is ``tighter'' when it is based on 
$V_{flat}$ as a rotation measure. The inset demonstrates the zoom-in to the infra-red bands. 
\label{fig_tght}}
\end{center}
\end{figure*}

\subsection {Fitting method}
The study of the statistical properties of the TFr requires establishing the slope 
and zero point of the relation. However, there is no general agreement which fitting
 method is best suited. For example, it was shown that
 the slope of the TFr is affected by a Malmquist bias (\citetalias{TC12}) which can be resolved 
 by applying an inverse least squares regression \citep{willick94}. Moreover, it is important to note
 that the vertical scatter of the TFr, which is crucial for the distance measure, is highly dependent 
on the slope. Thus, an intrinsically tight TFr may introduce a larger vertical scatter due to
a steeper slope \citep{v01}. 
As we are interested in the tightness of the TFr, while the Malmquist bias is minimal for our sample, we apply an
orthogonal regression where the best-fit model minimises the orthogonal distances 
from the points to the line. We apply a fitting method allowing for {\it bivariate correlated errors and an
intrinsic scatter} (BCES, \citealp{ akritas96}), using the python implementation developed by \citet{nemmen12}.
The main advantages of this method are that it takes errors in both directions into account,
it permits the measurement errors of both variables to be dependent (for example uncertainties due
to the inclination) and it assigns less weight to outliers and data points with large errors.

In order to accurately calculate the (intrinsic) scatter and tightness of the relations, the following 
measurement uncertainties were taken into account:
\begin{enumerate}
\item the errors in total magnitudes $M_{T}^{b},{i}(\lambda)$ due to the sky background,
distance uncertainties and the uncertainty in the photometric zero--point.
\item the errors on the rotational velocity measures $V_{max}$, $V_{flat}$ 
and $W_{50}^{i}$, see Table \ref{tbl_rot}.
\item the error on the kinematic inclination which affects both the internal extinction correction 
and the kinematic measure, introducing covariance in the errors.
\end{enumerate}

\subsection {Slope, scatter and tightness}
We measure the slope, scatter and tightness of the TFrs in 12 different bands with different 
kinematic measures, using the weighted orthogonal regression fit and taking correlated errors in 
both directions into account. It is important to point out that the 
comparisons are made for the samples with different numbers of galaxies:
32 for the UV and IR bands and 21 for the SDSS bands (see Section 3). 
However, we present the comparisons for the SDSS
subsample of 21 galaxies for all bands in Appendix A.

It has been suggested for some time that the slope of the TFr steepens from blue to red wavelengths \citep{aar79, tully82, v01}. 
We confirm this result by our study, which covers a much broader wavelength range. The variation of the
slope with passband is presented in Figure \ref{fig_slope}. Our result suggests that the slope 
as a function of wavelength in the mid--infrared bands stays more or less constant. Moreover, the TFr based on $V_{flat}$ is always showing the steepest slope in every passband (Figure \ref{fig_slope}). The steepest slope is found in the $K$ -band and is consistent with $-10$ mag or $4$ dex.

The vertical scatter ($\sigma$) in every passband was measured using each of the three velocity measures $W_{50}^{i}$, $V_{max}$ and $V_{flat}$.
The total observed scatter was calculated according to the following equation:
\begin{equation} 
\sigma = \sqrt{\frac{\chi^2}{N-1}},
\end{equation}
where $\chi^2$ is
\begin{equation*} 
\sum (M_{T}^{b,i} - (a\times log(V_{circ})+b))^2,
\end{equation*}
and where $V_{circ}$ stands for one of the three velocity measures 
$W_{50}^{i}$, $2V_{max}$ or $2V_{flat}$, {\it a} and {\it b} are the fitted 
slope and zero point of the relation respectively, and $N-1$ is the number of 
degrees of freedom. Errors on the scatter were estimated following a full error propagation calculation.
The vertical scatter in magnitudes, which is relevant for
distance measurements, is shown in Figure \ref{fig_scat} as a function of 
wavelength. It is clear from Figure \ref{fig_scat} that the vertical scatter is roughly constant in the
mid--IR bands, suggesting that there might be no preference for which mid--IR band to use 
as a distance indicator. Therefore, preference should be given to the one 
which suffers least from dust extinction and non--stellar contamination. However,
one can argue that the $r$--band TFr based on $W_{50}^{i}$ should be used as a distance estimation tool, 
since it demonstrates the smallest vertical scatter. Interestingly, \citet{v01} had found a very similar result. 
Yet, it is important to keep in mind that the vertical scatter
is a slope--dependent measure, and an intrinsically tight TFr will demonstrate a large vertical scatter
if the slope of the relation is steep. Moreover, it is remarkable that for the $UV$ and optical bands ($FUV$ to $z$) the vertical scatter 
may be smaller when the relation is based on $W_{50}^{i}$, with the smallest scatter of $\sigma = 0.23$ mag in the $r$ band. 
However, for the redder bands ($J$ to $4.5 \mu m$) the smallest vertical scatter $\sigma = 0.33$ mag is found in the
 3.6 $\mu$m band TFr based on $2V_{flat}$. This is due to the fact that when the relation is based on $2V_{flat}$, the slope steepens 
more significantly for the $UV$ and optical bands than for the infrared. 

\begin{figure*}
\begin{center}
\includegraphics[scale=0.60]{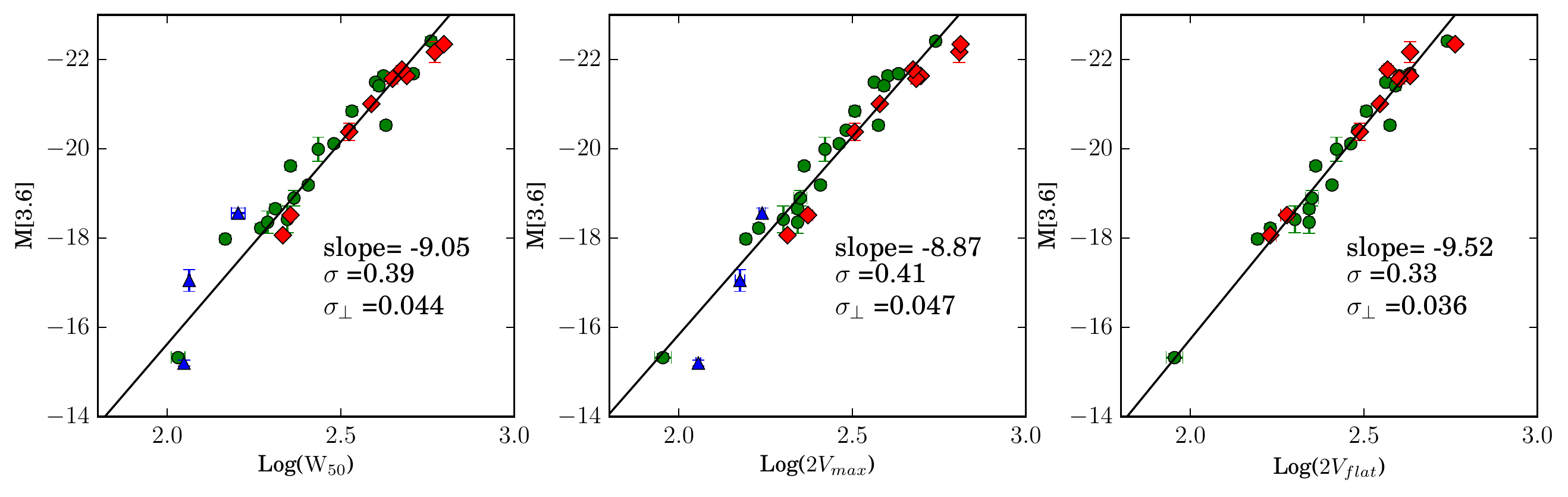}
\caption{
The 3.6 $\mu$m TFrs based on the different kinematic measures $W_{50}^{i}$ -- left, 
$V_{max}$ -- middle and $V_{flat}$ -- right. Green symbols 
show flat rotation curves ($V_{max} = V_{flat}$), and red symbols indicate 
galaxies with declining rotation curves ($V_{max} > V_{flat}$). Blue symbols 
indicate galaxies with rising rotation curves ($V_{max} < V_{flat}$). These galaxies 
were not included when fitting the model. 
\label{fig_36tf}}
\end{center}
\end{figure*}

The tightness ($\sigma_{\perp}$) of the TFr is the perpendicular scatter between the data points and the linear model, 
which provides information on how ``tight'' the data
are spread around the regression line. It is slope independent and has been recently used for testing galaxy formation and evolution models \citep{manolistfr}.
Therefore, the tightness provides important information on the intrinsic properties of the TFr.
We calculate tightness using the following 
formula:
\begin{equation} 
\sigma_{\perp} = \sqrt{\frac{\sum d_{i}^2}{N-2}},
\end{equation} 
where $N-2$ is the amount of degrees of 
freedom and $d_{i}$ is the perpendicular distance of each point to a model line:

\begin{equation*} 
d_{i}=\sqrt{\bigg(\frac{x_{i}+ay_{i}-ab}{a^{2}+1}-x_{i}\bigg)^{2}+\bigg(a \times \frac{
x_{i}+ay_{i}-ab}{a^{2}+1}+b-y_{i}\bigg)^{2}},
\end{equation*} 
here $x_{i}$ and $y_{i}$ are the coordinates of each measured point, in our 
case $log(L_{\lambda}/L_{\odot})$ and $log(V_{circ})$ respectively, while {$a$} and {$ b$} are the 
slope and the zero point of a model line. Errors on the tightness were estimated following a full error
propagation calculation. The tightness of the TFr as a function of wavelength is shown in Figure \ref{fig_tght}.
It is also roughly constant in the mid--IR bands with the tightest correlation at 3.6 $\mu$m equal 
to  $\sigma_{\perp} = 0.036$ dex. 
Moreover, independently of the photometric band, the TFrs tend to be somewhat tighter when based on $2V_{flat}$ (Figure \ref{fig_tght}).
The values of the statistical properties of the TFrs (slope, zero point, scatter and tightness) can be found in Tables \ref{tbl_slope}, \ref{tbl_scatter} and  \ref{tbl_tght}. 


\subsection{A closer look at the 3.6 $\mu$m TFr}

Figure \ref{fig_36tf} shows the TFr in the 3.6 $\mu$m band based on $W_{50}^{i}$, $V_{max}$ and $V_{flat}$. 
According to our fit, the $M_{[3.6]}^{T,b,i}$--$V_{flat}$ correlation can be described as : 
\begin{equation} 
M_{[3.6]}^{T,b,i} = (-9.52\pm0.32)\times log(2V_{flat})+3.3\pm0.8
\label{eq_mag}
\end{equation} 
and the $L_{[3.6]}^{T,b,i}(L_{\odot})$--$V_{flat}$ correlation as
\begin{equation} 
log(L_{[3.6]}^{T,b,i}) = (3.7\pm0.11)\times log(2V_{flat})+1.3\pm0.3,
\label{eq_lum}
\end{equation} 
based on $M_{\odot} (3.6 \mu m)=3.24$ mag \citep{oh08}. Here $M_{[3.6]}^{T,b,i}$ is the total magnitude,
corrected for Galactic and internal extinction,
$L_{[3.6]}^{T,b,i}$ is the luminosity, presented in solar luminosities, and $V_{flat}$ is the rotational velocity
of the flat part of the extended H{\sc i} rotation curve in $km/s$.

 Eqn. \ref{eq_mag} and  Eqn. \ref{eq_lum} describe the tightest of the TFrs, with an observed tightness equal to
$\sigma_{\perp,obs}= 0.036 \pm 0.010$ dex. Without considering the observational errors, 
$\sigma_{\perp,obs}$ presents an upper limit on the intrinsic tightness 
of the TFr$_{[3.6]-V_{flat}}$ of $\sigma_{\perp,int} < \sigma_{\perp,obs}=0.036$ dex.
This is $0.02$ dex smaller than the observed tightness of the Baryonic TFr for 
gas--rich galaxies, found by \citet{manolistfr}, using $W_{50}^{i}$ as a rotational velocity measure.

\begin{figure}
\begin{center}
\includegraphics[scale=0.90]{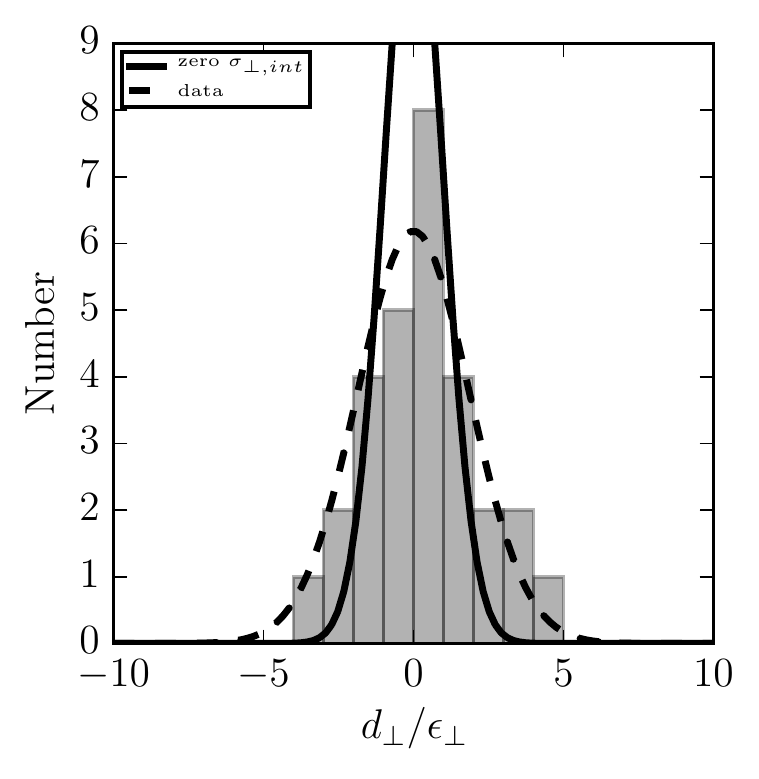}
\caption{Histogram of the perpendicular distances from the data points to the line ($d_{\perp,i}$) in the $L_{[3.6]}^{T,b,i}(L_{\sun})$--$V_{flat}$
relation, normalised by the perpendicular errors $\epsilon_{\perp,i}$. The standard normal distribution, which would 
be expected for a zero intrinsic tightness is shown with a black line. The best--fit to the data, weighted by the Poisson errors, 
is shown with the dashed line with a standard deviation of 1.87 $\pm$ 0.13.
\label{fig_norm}}
\end{center}
\end{figure}

\begin{figure}
\begin{center}
\includegraphics[scale=0.45]{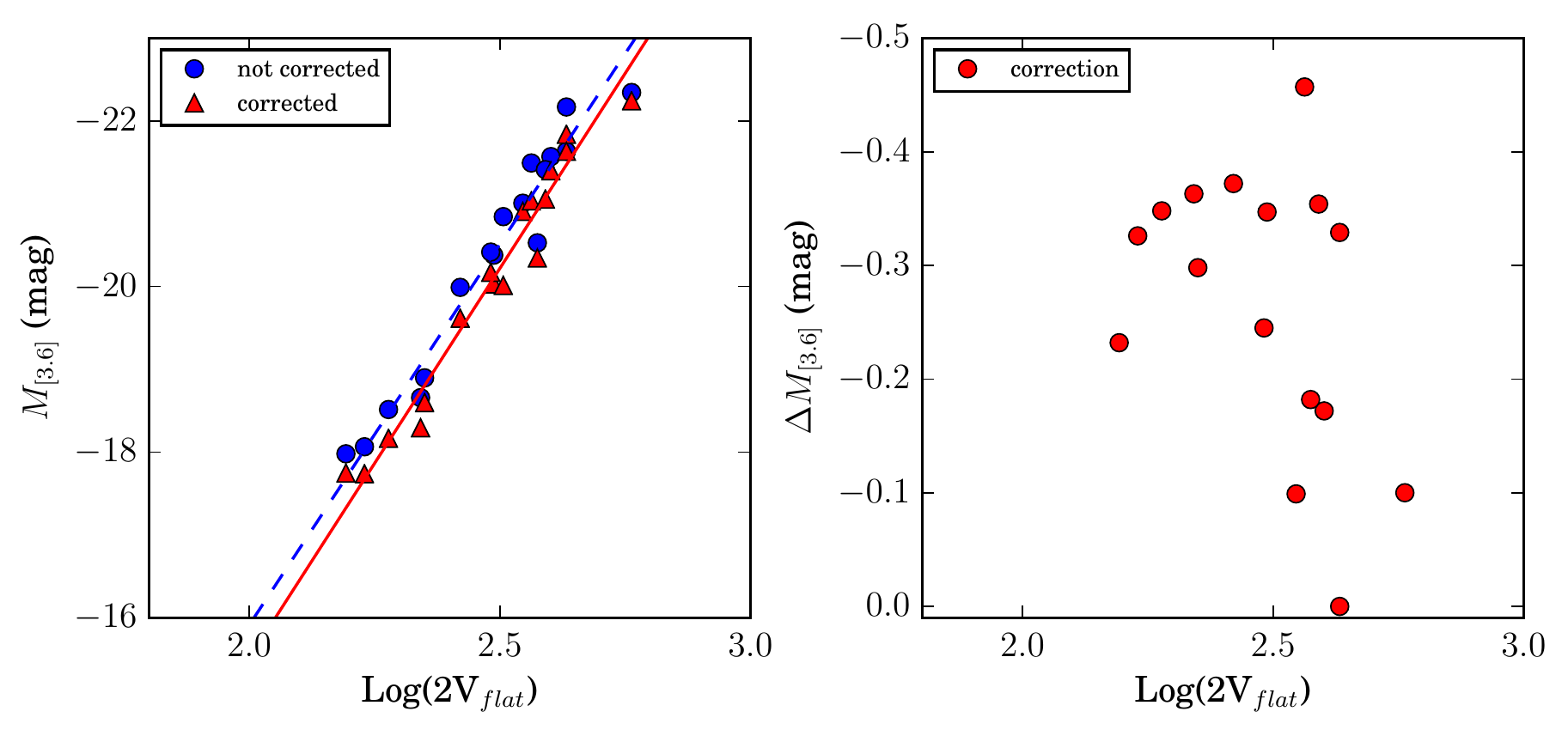}
\caption{Left panel: The  $M_{[3.6]}^{T,b,i}$--$V_{flat}$ relation is shown with blue points. The relation using 
total magnitudes corrected for non--stellar contamination  is shown with red points. Right panel: The correction
for non--stellar contamination as a function of rotational velocity $2V_{flat}$.
\label{fig_oldsts}}
\end{center}
\end{figure}

\begin{figure*}
\begin{center}
\includegraphics[scale=0.60]{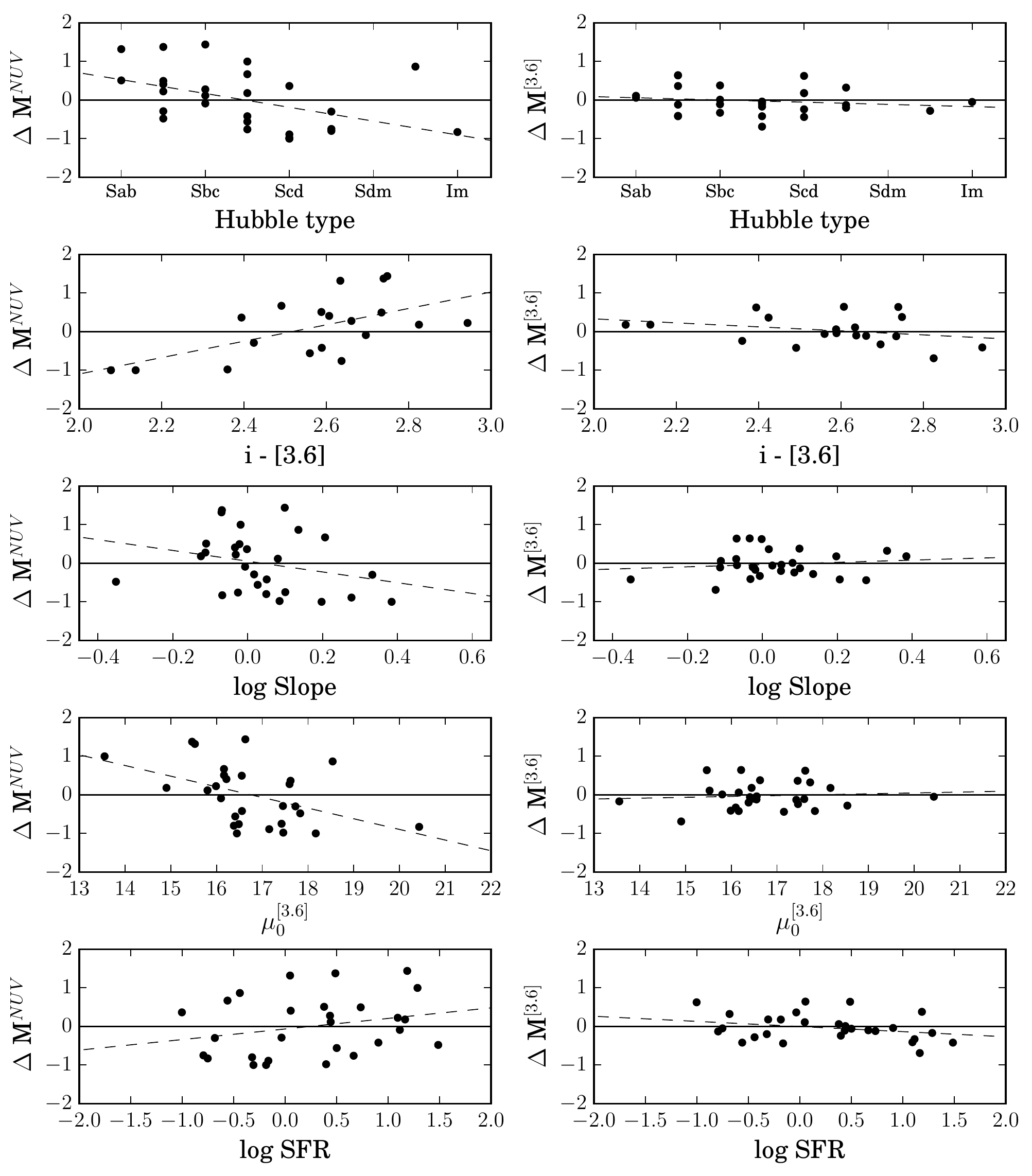}
\caption{Residuals of the $M_{\lambda}-V_{flat}$ TFrs in the NUV and 3.6 $\mu$m bands as a function of global galactic properties.   
\label{fig_res1}}
\end{center}
\end{figure*}

\begin{figure}
\begin{center}
\includegraphics[scale=0.70]{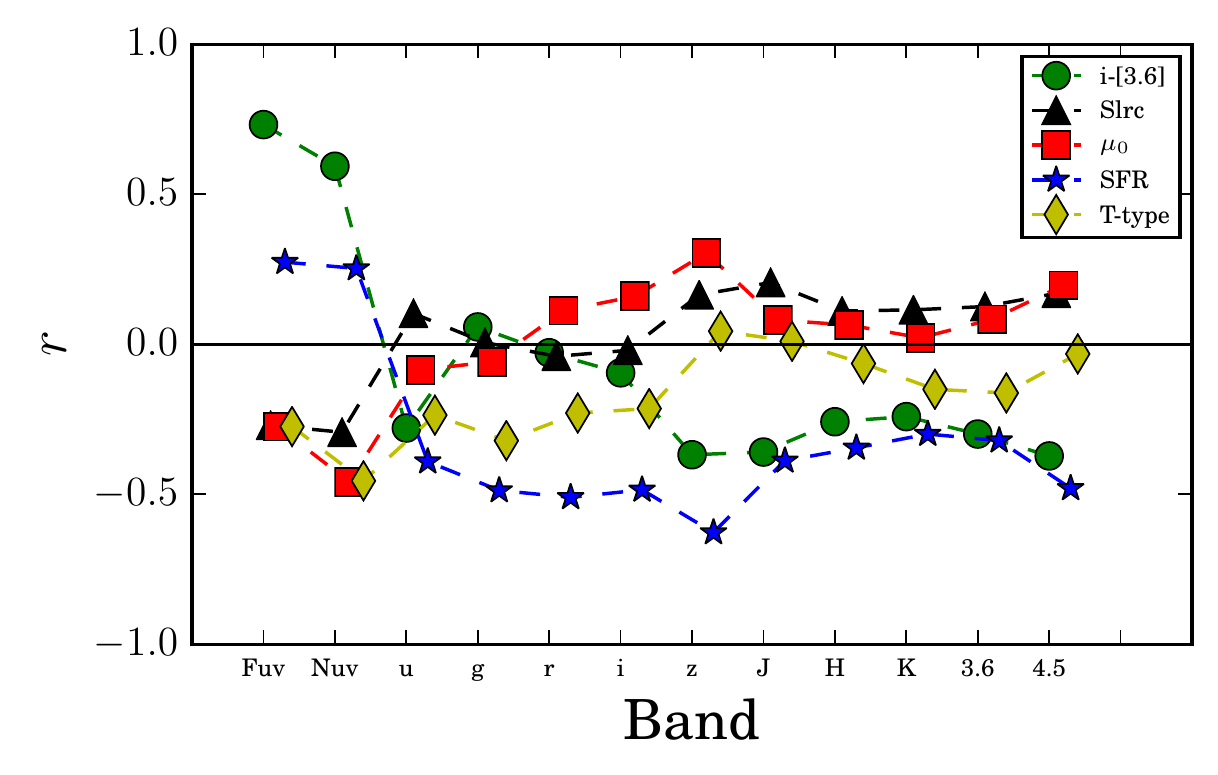}
\caption{Pearson's correlation coefficients between $M_{T}(\lambda)$-- log($2V_{flat}$) and 
global galactic properties as a function of wavelength.   
\label{fig_ro}}
\end{center}
\end{figure}

Further, we can estimate the intrinsic tightness of the TFr$_{[3.6],V_{flat}}$ by comparing the perpendicular distance from each data point to the model line $d_{\perp, i}$ taking into account the measurement error $\epsilon_{\perp, i}$, where  $\epsilon_{\perp, i}$ is based on the observational errors on the luminosity ($\epsilon_{L_{[3.6],i}}$) and on the rotational velocity ($\epsilon_{V_{flat},i}$) of each data point, projected onto the direction perpendicular to the model line:
\begin{equation} 
\epsilon_{i}^2=(\epsilon_{log(L_{[3.6]}),i} \times \frac{1}{\sqrt{1+a^{2}}})^{2}+(\epsilon_{log(2V_{flat}),i} \times \frac{a}{\sqrt{1+a^{2}}})^{2},
\label{eq_er}
\end{equation} 
where $a$ is the slope of the line. The uncertainty on to the distance measurement is included in the error
on the total luminosity. For more details on the derivation of Eqn. \ref{eq_er} see \citealp{manolistfr} (Appendix B). 
If the observed tightness of the relation would be only due to the measurement errors, the histogram
of $d_{\perp, i}/\epsilon_{\perp, i}$ would then follow a standard normal distribution.
Yet, it is clear from Figure \ref{fig_norm} that the spread of  $d_{\perp, i}/\epsilon_{\perp, i}$ is larger than
expected from a standard normal distribution, with a measured standard deviation of 1.87 $\pm$ 0.13 (dashed curve). Therefore, a small but non--zero intrinsic perpendicular 
scatter ($\sigma_{\perp}$) is present in the TFr$_{[3.6]-V_{flat}}$.

To obtain the best estimate for the intrinsic tightness, we present it as follows :
\begin{equation} 
\sigma_{\perp,int}=\sqrt{\sigma_{\perp,obs}^{2}-\sigma_{\perp,err}^{2}},
\end{equation} 
where $\sigma_{\perp,err}=0.025$ dex is the perpendicular scatter due to the measurement errors only. 
Thus, we obtain an estimate for the intrinsic perpendicular scatter $\sigma_{\perp,int} \sim 0.026$ dex.
It is important to keep in mind that this result depends on how accurately the observational errors
can be determined. Therefore, any underestimate of the observational errors will lead to a decrease of the intrinsic 
tightness and vice versa. 

As was already mentioned before, previous studies \citep{ meidt12, miguel15} concluded that Spitzer 3.6 $\mu$m
luminosities represent not only the old stellar population. Up to 30\% of the 3.6 $\mu$m light might be coming from warm dust
which is heated by young stars and re--emitted at longer wavelengths. Moreover, asymptotic giant branch (AGB) stars appear to peak at 3.6 $\mu$m as well. 
To test the effect of contamination by dust and/or AGB stars on the tightness of the TFr, we constructed a subsample of 18 galaxies which were studied as part of the S$^4$G analysis
 by \citet{miguel15}. In this study, the 3.6 $\mu$m images were decomposed into stellar and non-stellar contributions using an Independent Component Analysis described in \citet{meidt12}.
We compared the statistical properties of the TFrs in the observed 3.6 $\mu$m band and in the 3.6 $\mu$m band corrected for non--stellar contamination, as demonstrated in Figure \ref{fig_oldsts} (left panel). 
The results of the comparison can be found in Table \ref{tbl_correct}. It is clear that the scatter and tightness of the 
TFr at 3.6 $\mu$m can be slightly reduced if the non--stellar contamination is corrected for, especially when the TFr is based on
$V_{flat}$ as a rotational velocity measure. The very weak correlation between the correction for non--stellar contamination 
and rotational velocity (Figure \ref{fig_oldsts}, right panel) results in a slight steepening of the TFr slope. 
However, the difference in the scatter ($\Delta \sigma=0.03$ mag) and tightness ($\Delta \sigma_{\perp}=0.004$ dex) is too small ($\sim$ 10\%) to draw definite conclusions. Hence, a more detailed study of this subject should be done with a larger sample of galaxies for which a decomposition into stellar and non--stellar emission has been performed.

\begin{table}
\begin{tabular}{lllllll}
\hline
 N=19 & \multicolumn{3}{c}{Observed}& \multicolumn{3}{c}{Corrected}\\
               &$W_{50}^{i}$& $V_{max}$ &$V_{flat}$ &$W_{50}^{i}$& $V_{max}$ &$V_{flat}$\\
\hline                
slope (mag)           &-8.52&-8.56&-9.20&-8.77&-8.77&-9.47\\
$\sigma$ (mag)        &0.40&0.45&0.32&0.38&0.41&0.29\\
$\sigma_{\perp}$ (dex)&0.043&0.046&0.036&0.043&0.046&0.031\\
\hline
\end{tabular}
\caption{Slope, scatter and tightness of the TFrs, constructed for 19 galaxies in the observed 3.6 $\mu$m band and in the 3.6 $\mu$m band, corrected for the non--stellar contamination.
Column(1): name of the parameter;
Column (2-4): slope, scatter and tightness of the TFrs in the observed 3.6 $\mu$m band (based on different velocity measures);
Column (5-7): slope, scatter and tightness of the TFrs in the 3.6 $\mu$m band, corrected for non--stellar emission (based on different velocity measures);
}
\label{tbl_correct}
\end{table}   

\subsection {Need for a $2^{nd}$ parameter?} 
For many decades it was suggested that the scatter in the TFr can be reduced by adding 
a second parameter such as colour, morphological type or velocity dispersion \citep{rubin85, tp00, cortese14}. 
This parameter is usually derived from the correlations of the 
residuals of the TFr with global galactic properties. It has been shown that the residuals of the TFr based on $W_{50}^{i}$
correlate well with the colour or morphological type of galaxies \citep{aar84, rubin85, russell04}. 
The correlations are usually found in the blue bands which 
tend to have much larger scatter, and found to be completely absent in the red 
bands where the scatter is already very small \citep{tp00,v01}. 
However,  \citet{sorce12} found a colour term present in the residuals of the 3.6 $\mu$m TFr, 
which allowed them to reduce the observed scatter by 0.05 mag. 

We examine the residuals $\Delta M_{\lambda}$ of the TFrs in each band based on $V_{flat}$, and 
investigate possible correlations with global galactic properties such as star formation rate (SFR), central surface brightness, and the outer slope of the rotation curve. First, we consider in detail two extremes $\Delta M_{NUV}$ and $\Delta M_{[3.6]}$ based on $V_{flat}$.
From Figure \ref{fig_res1} it is clear that while NUV residuals show hints for correlations with all 
galactic parameters, these hints completely disappear for the 3.6 $\mu$m residuals. 

To quantitatively describe the strengths of the correlations between TFr residuals and global galactic properties, we 
calculate Pearson's coefficients $r$, a measure of the linear correlation between two variables.
Figure \ref{fig_ro} shows the Pearson's coefficients  $r$ as a function of wavelength for the correlations between $\Delta M^{\lambda}$ and
various galactic properties such as $i-[3.6]$ colour, the outer slope of the rotation curve (see Section 5.2), central 
 surface brightness at 3.6 $\mu$m, star formation rate and morphological type. It is clear from Figure \ref{fig_ro} that the Pearson's coefficients  do not suggest any strong correlations between residuals of the TFrs and various galactic properties in any band, except for the $FUV$ where a prominent correlation with the $i-[3.6]$ colour is found. This result is in agreement with previous studies for blue bands. However, there is no evidence for a significant correlation between $\Delta M_{[3.6]}$ and the $i-[3.6]$ colour ($r = -0.29$), despite the previous suggestions by \citet{sorce12}. Nonetheless, the strength for the correlation between $\Delta M_{[3.6]}$ and the $i-[3.6]$ colour was not presented in the \citet{sorce12} study, therefore we can not 
perform a quantitative comparison. In conclusion, we do not find any second parameter, which would help to reduce the scatter in the near--infrared TFr.

 \section{Summary and Conclusions}
In this paper, we present an empirical study of the multi--wavelength Tully--Fisher relation, taking advantage of spatially resolved H{\sc i} kinematics. 
This study aims to investigate the statistical properties of the TFrs in 12 photometric bands, using three rotational velocity measures: $W_{50}^{i}$ from the global H{\sc i} profile, and $V_{max}$ and $V_{flat}$ from high--quality, spatially--resolved H{\sc i} rotation curves.
The galaxies in our sample were selected to have independently measured Cepheid or/and TRGB distances. This allowed us to calibrate
the TFr with minor distance uncertainties ($\sigma_{dist}=0.07$ mag, instead of $\sigma_{dist}=0.41$ mag when using Hubble flow distances). 

First, we present a slope--independent perpendicular scatter ($\sigma_{\perp}$) of the TFr, 
which describes how tight the data points are spread around the model line. 
We study the tightness as a function of wavelength for TFrs based on
different rotational velocity measures (Section 6.2). We find that the tightness $\sigma_{\perp}$ of the TFr
improves significantly from the blue to the infrared bands, but it levels off for the near--infrared bands, 
with the largest $\sigma_{\perp} = 0.043$ dex in the H--band and 
the smallest $\sigma_{\perp} =0.036$ dex in the 3.6 $\mu$m band, using $V_{flat}$ as a 
rotational velocity measure. We find that the latter is not consistent with a zero intrinsic perpendicular scatter 
indicating that the measured $\sigma_{\perp,obs}$ can not be completely explained by the measurement errors (see Section 6.3). 
Nevertheless, the TFr based on the 3.6 $\mu$m luminosities and $V_{flat}$ provides the tightest constraint on theories of
galaxy formation and evolution. Indeed, such a tight correlation between the 3.6 $\mu$m luminosity of a galaxy with the velocity of the outer
most point of the rotation curve suggests an extremely tight correlation between the mass
of the dark matter halo and its baryonic content. Certainly, 3.6 $\mu$m light has been 
considered as the best tracer of the total stellar mass of 
galaxies which dominates the baryonic mass. However, many observational studies have shown 
that not only old stars, but also hot dust and AGB stars might contribute to
this light, up to 30\% in some cases. We have shown that the observed tightness and scatter 
of the TFr can be somewhat reduced if the 3.6 $\mu$m light is corrected for non--stellar contamination 
(see Section 6.3). More studies should be done to further investigate this 
effect, using a larger sample of galaxies for which the decomposition of the light into old stars 
and contamination can be performed. 

An obvious next step in studying the tightness of the TFr is to measure the slope, scatter and 
tightness of the baryonic TFr (BTFr) and compare this with
measurements derived for the 3.6 $\mu$m band. This approach introduces more
uncertainties related to estimating the stellar mass. For instance, \citet{manolistfr} found 
a larger perpendicular scatter of the BTFr, even though they considered a sample of heavily 
gas--dominated galaxies for which uncertainties in stellar mass are less significant. That study was done using only $W_{50}^{i}$ as a rotational velocity measure. A forthcoming paper will discuss the statistical properties of the BTFr with resolved H{\sc i} kinematics. 
 
Next, we study the slope of the TFr as a function of wavelength, using $W_{50}^{i}$,
 $V_{max}$ and $V_{flat}$ (see Section 6.2). We confirm the results of previous 
 studies \citep{aar79, tully82, v01}, that the slope of the TFr steepens toward longer wavelengths by 
broadening the study over a wider wavelength range. 
The steepening of the slope results from the fact that redder galaxies are much brighter than bluer galaxies
at longer wavelengths. Galaxies that are bright in the infrared tend to rotate more rapidly. 
Therefore, at longer wavelengths  the high--mass end of the TFr will rise faster than the low--mass end.
 In addition, we find that the TFr based on $V_{flat}$ as a rotational velocity
 measure has the steepest slope in every photometric band. Massive galaxies tend to have declining 
 rotation curves with $V_{max} > V_{flat}$ (see Section 4.1, Figure \ref{fig_vres}). If $V_{flat}$ is used 
 as a rotation velocity measure, bright galaxies have lower velocities than when measured with $W_{50}^{i}$ and/or $V_{max}$. This
 difference reduces the velocity range over which the galaxies are distributed. 
This effect steepens the slope of the TFr as well. Similar results were found by \citet{v01} and \citet{noord07}.
 Moreover, the use of $V_{flat}$ ``straightens" the TFr and removes a possible curvature in the TFr
 at the high--mass end \citep{neill14,noordver07}.

Subsequently, we discussed the vertical scatter ($\sigma$) of the TFr as a function of wavelength, using three
rotational velocity measures (see Section 6.2). It is well known, that the vertical scatter of the TFr
is strongly dependent on the slope. Thus, even an intrinsically tight
correlation can be found to have a large vertical scatter if the slope is steep. The vertical scatter
of the TFr is mostly discussed in the context of determining distances to galaxies.  
We find the smallest vertical scatter in the $r$--band, using $W_{50}^{i}$ as a rotational velocity measure,
confirming the result found by \citet{v01}. Moreover, we find that the vertical scatter in the
3.6 $\mu$m band ($\sigma=0.39$mag) to be lower than previously reported by \citet{sorce12} 
for the 3.6 $\mu$m  band ($\sigma=0.44$mag), and by \citet{neill14} for the 3.4 $\mu$m band ($\sigma=0.54$ mag).
These comparisons are done using $W_{50}^{i}$ as a rotational velocity measure. 
Besides, we find $\sigma$ to be smaller when using $W_{50}^{i}$ as a velocity measure for the $FUV$ and optical bands 
($FUV-z$). For the infrared bands (J to 4.5$\mu$m), $\sigma$ is smaller when the TFr is based on $V_{flat}$.
This result suggests that $\sigma$ in the infrared bands is less sensitive to the slope steepening with $V_{flat}$. 

We searched for a second parameter that can possibly help to reduce the vertical scatter
of the TFr. We considered the residuals of the TFrs ($\Delta M_{\lambda}$--$V_{flat}$ ) in every band (see Section 6.4)
and find no significant correlations between the residuals of the TFrs and main 
galactic properties (SFR, central surface brightness, outer slope of the 
rotation curve, morphological type and $i-[3.6]$ colour, see Figure 
\ref{fig_ro}). 
Even though the $UV$ bands show hints for correlations between the residuals and some of the global properties such as SFR (see Figure \ref{fig_res1}), no correlations are found in the red bands. This suggests that these correlations are triggered by different stellar populations
in early--type and late--type galaxies of the same UV luminosity and not by the difference in $V_{flat}$ governed by the dark matter halo.
Lastly, it is important to mention that we do not find any correlation between the TFr residuals $\Delta M_{[3.6]}$--$W_{50}^{i}$ and the colour term $i-[3.6]$ (Pearson's coefficient $r=0.1$), contrary to the result reported previously by \citet{sorce12}.

As was shown by \citet{sorce16}, the size of the sample may have a significant impact on the scatter of the TFr. 
Therefore, it is necessary to point out that the limited size of our sample might contribute to the uncertainties in the slope,
scatter and zero point of the TFrs.
However, it is very expensive to establish a large sample of spiral galaxies which have both independently measured distances 
and resolved H{\sc i} kinematics. Nonetheless, this challenge will be possible to meet with the H{\sc i} imaging surveys 
that are planned for new observational facilities, such as Apertif \citep{aperitif}, MeerKAT \citep{meerkat} and ASKAP \citep{askap},
providing resolved H{\sc i} kinematics for many thousands of galaxies.

\section*{acknowledgements}
AP is grateful to Emmanouil Papastergis for fruitful discussions and useful comments.  
AB acknowledges financial support from the CNES (Centre National d'Etudes Spatial -- France).
We thank the staff of the GMRT who made our observations possible. 
The GMRT is run by the National Centre for Radio Astrophysics of the 
Tata Institute of Fundamental Research. This
research has made use of the NASA/IPAC Extragalactic Database (NED)
which is operated by the Jet Propulsion Laboratory, California Institute
of Technology, under contract with the National Aeronautics and Space
Administration.  This research made use of Montage, funded by the National 
Aeronautics and Space Administration's Earth Science Technology Office, 
Computational Technologies Project, under Cooperative Agreement Number
 NCC5-626 between NASA and the California Institute of Technology. 
The code is maintained by the NASA/IPAC Infrared Science Archive.
We acknowledge financial support from the DAGAL network
from the People Programme (Marie Curie Actions) of the European Union's
Seventh Framework Programme FP7/2007-2013/ under REA grant agreement
number PITNGA-2011-289313. We acknowledge the Leids Kerkhoven--Bosscha
Fonds (LKBF) for travel support.

\bibliographystyle{mnras.bst} 
\bibliography{Chap3} 

\appendix 
\section{TFrs for the SDSS subsample}
\begin{figure}
\includegraphics[scale=0.60]{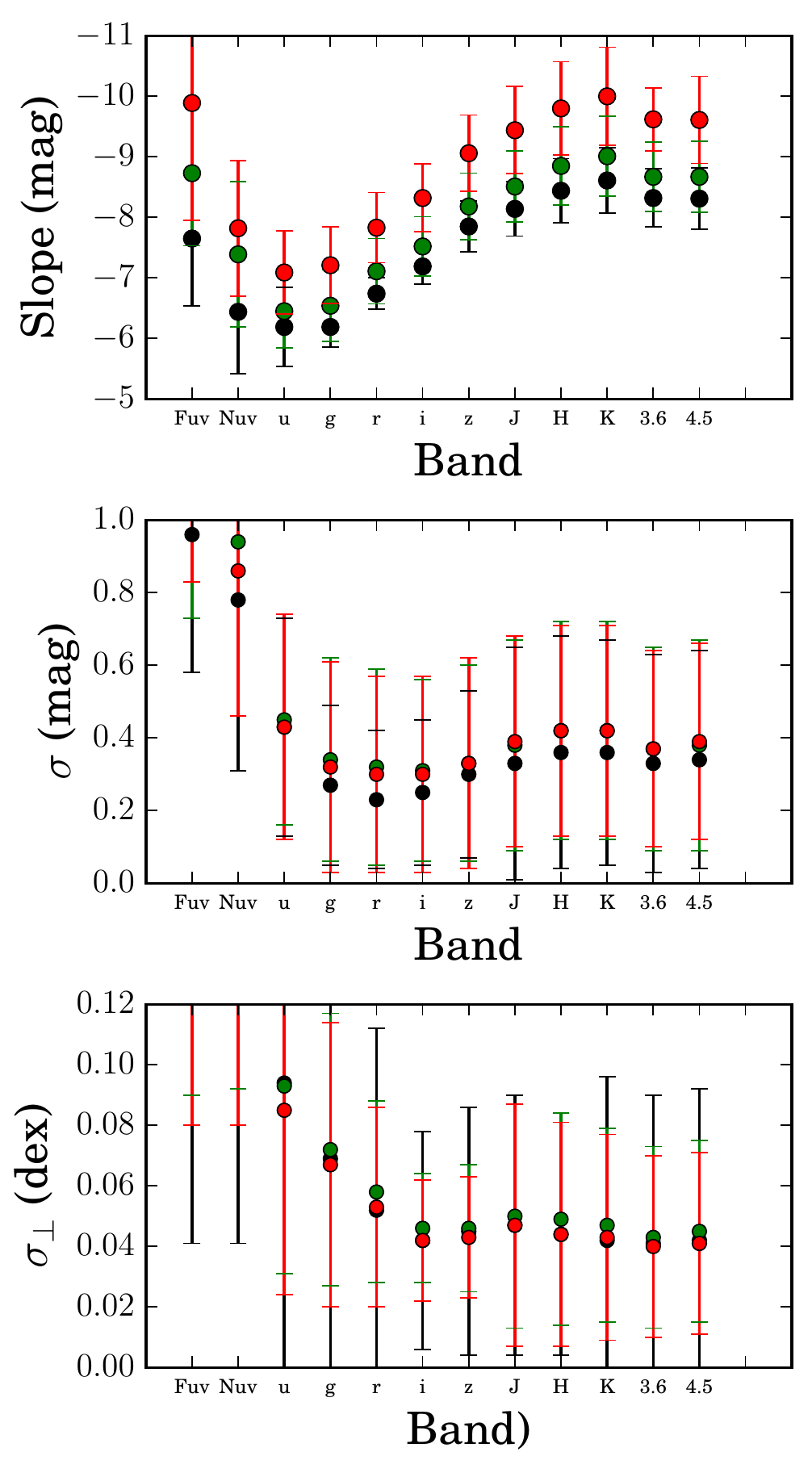}
\caption{Slope, scatter and tightness of the TFr of the SDSS subsample as a function of wavelength, calculated using 
different rotation measures. With black points indicated values measured for 
the TFr based on $W_{50}^{i}$, with green based on $V_{max}$ and with red based on $V_{flat}$. 
\label{fig_slred}}
\end{figure}
In this Appendix we present the results for the smaller SDSS sample. In this subsample we consider only 21 galaxies 
which have photometry from all 12 bands. 
This allows us to compare the slope, scatter and tightness of the TFrs at various wavelengths for the same number of galaxies. 
We recall that the lack of SDSS data for our full sample resulted in this smaller SDSS subsample. 
Figure \ref{fig_slred} demonstrates the slope, scatter and tightness of the TFrs for this subsample. 
Even though the number of galaxies is smaller, there is no significant difference in trends compared to results based on the full sample.
Table \ref{tbl_red} summarises the measurements for the reduced TFr sample. 
\begin{table*}
\begin{tabular}{llllllllll}
\hline
 Band&\multicolumn{3}{c}{Slope (Mag)}&\multicolumn{3}{c}{Scatter (Mag)}&\multicolumn{3}{c}{Tightness $\sigma_{\perp}$ (dex)}\\
 &$W_{50}^{i}$& $V_{max}$ &$V_{flat}$ &$W_{50}^{i}$& $V_{max}$ &$V_{flat}$&$W_{50}^{i}$& $V_{max}$ &$V_{flat}$\\
 \hline
 FUV &-7.65$\pm$1.11& -8.73$\pm$1.20& -9.89$\pm$1.94 & 0.96$\pm$0.38&1.13$\pm$0.40&1.17$\pm$0.34& 0.132$\pm$0.091&0.135$\pm$0.045&0.125$\pm$0.045 \\    
 NUV &  -6.44$\pm$1.02 & -7.39$\pm$1.20 & -7.82$\pm$1.12 &  0.78$\pm$0.47 & 0.94$\pm$0.48 & 0.86$\pm$0.40  & 0.138$\pm$0.097& 0.141$\pm$0.049& 0.129$\pm$0.049 \\    
 u   &  -6.19$\pm$0.65 & -6.45$\pm$0.61 & -7.09$\pm$0.69 &  0.43$\pm$0.30 & 0.45$\pm$0.29 & 0.43$\pm$0.31  & 0.094$\pm$0.124& 0.093$\pm$0.062& 0.085$\pm$0.061 \\   
 g   &  -6.19$\pm$0.33 & -6.54$\pm$0.59 & -7.21$\pm$0.63 &  0.27$\pm$0.22 & 0.34$\pm$0.28 & 0.32$\pm$0.29  & 0.069$\pm$0.089& 0.072$\pm$0.045& 0.067$\pm$0.047 \\   
 r   &  -6.74$\pm$0.26 & -7.11$\pm$0.54 & -7.83$\pm$0.58 &  0.23$\pm$0.19 & 0.32$\pm$0.27 & 0.30$\pm$0.27  & 0.052$\pm$0.060& 0.058$\pm$0.030& 0.053$\pm$0.033 \\   
 i   &  -7.19$\pm$0.29 & -7.52$\pm$0.49 & -8.32$\pm$0.56 &  0.25$\pm$0.20 & 0.31$\pm$0.25 & 0.30$\pm$0.27  & 0.042$\pm$0.036& 0.046$\pm$0.018& 0.042$\pm$0.020 \\   
 z   &  -7.85$\pm$0.42 & -8.18$\pm$0.55 & -9.06$\pm$0.63 &  0.30$\pm$0.23 & 0.33$\pm$0.27 & 0.33$\pm$0.29  & 0.045$\pm$0.041& 0.046$\pm$0.021& 0.043$\pm$0.020 \\   
 J   &  -8.14$\pm$0.45 & -8.51$\pm$0.59 & -9.44$\pm$0.72 &  0.33$\pm$0.32 & 0.38$\pm$0.29 & 0.39$\pm$0.29  & 0.047$\pm$0.043& 0.050$\pm$0.037& 0.047$\pm$0.040 \\   
 H   &  -8.44$\pm$0.53 & -8.85$\pm$0.65 & -9.80$\pm$0.77 &  0.36$\pm$0.32 & 0.42$\pm$0.30 & 0.42$\pm$0.29  & 0.044$\pm$0.040& 0.049$\pm$0.035& 0.044$\pm$0.037 \\   
 K   &  -8.61$\pm$0.54 & -9.01$\pm$0.66 & -10.0$\pm$0.81 &  0.36$\pm$0.31 & 0.42$\pm$0.30 & 0.42$\pm$0.29  & 0.042$\pm$0.054& 0.047$\pm$0.032& 0.043$\pm$0.034 \\   
3.6  &  -8.32$\pm$0.48 & -8.67$\pm$0.57 & -9.62$\pm$0.52 &  0.33$\pm$0.30 & 0.37$\pm$0.28 & 0.37$\pm$0.27  & 0.041$\pm$0.049& 0.043$\pm$0.030& 0.040$\pm$0.030 \\   
4.5  &  -8.31$\pm$0.51 & -8.67$\pm$0.59 & -9.61$\pm$0.72 &  0.34$\pm$0.30 & 0.38$\pm$0.29 & 0.39$\pm$0.27  & 0.042$\pm$0.050& 0.045$\pm$0.030& 0.041$\pm$0.030 \\    
\hline
\end{tabular}
\caption{The slope, scatter an tightness for the reduced sample. 
Column (1): band;
Column (2)-Column (4): slopes of the TFrs based on $W_{50}^{i}$, $V_{max}$ and $V_{flat}$, measured in magnitudes;
Column (5)-Column(7): scatter of the TFrs based on $W_{50}^{i}$, $V_{max}$ and $V_{flat}$, measured in magnitudes;
Column (8)-Column (10): tightness of the TFrs based on $W_{50}^{i}$, $V_{max}$ and $V_{flat}$, measured in dex;
}
\label{tbl_red}
\end{table*}            

\section{TFrs, using $2.2h$ magnitudes}
\begin{figure}
\includegraphics[scale=0.60]{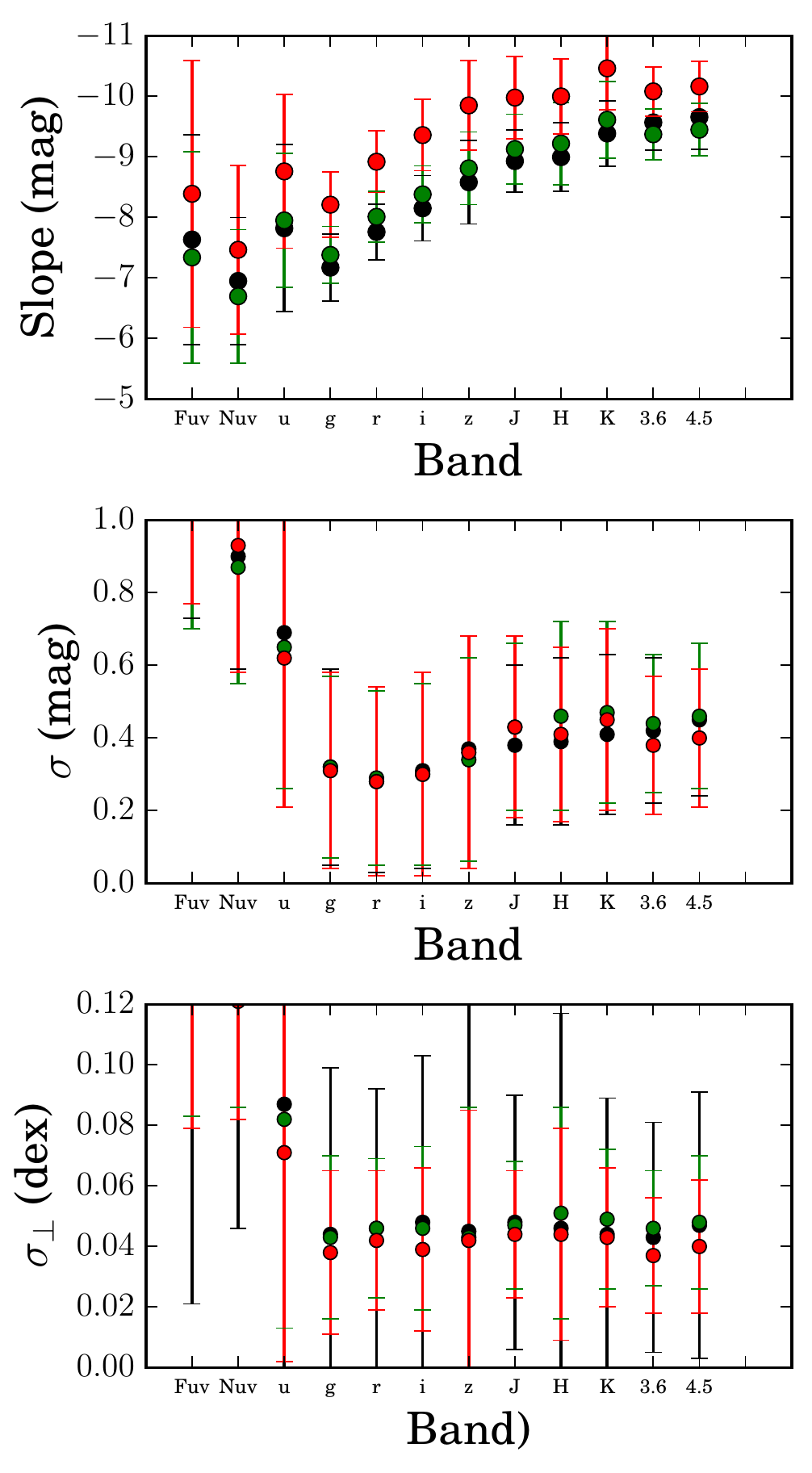}
\caption{Slope, scatter and tightness of the TFr based on magnitudes measured at $2.2h$, as a function of wavelength, calculated using 
different rotation measures. With black points indicated values measured for 
the TFr based on $W_{50}^{i}$, with green based on $V_{max}$ and with red based on $V_{flat}$.  
\label{fig_sl22}}
\end{figure}
\begin{table*}
\begin{tabular}{llllllllll}
\hline
 Band&\multicolumn{3}{c}{Slope (Mag)}&\multicolumn{3}{c}{Scatter (Mag)}&\multicolumn{3}{c}{Tightness $\sigma_{\perp}$ (dex)}\\
 &$W_{50}^{i}$& $V_{max}$ &$V_{flat}$ &$W_{50}^{i}$& $V_{max}$ &$V_{flat}$&$W_{50}^{i}$& $V_{max}$ &$V_{flat}$\\
 \hline
 FUV &  -7.63$\pm$1.73 & -7.33$\pm$1.74 & -8.39$\pm$2.20 &  1.13$\pm$0.40  &1.10$\pm$0.40 &  1.21$\pm$0.44  & 0.143 $\pm$ 0.122& 0.144 $\pm$ 0.061 & 0.140 $\pm$ 0.061  \\    
 NUV &  -6.95$\pm$1.04 & -6.69$\pm$1.10 & -7.46$\pm$1.38 &  0.90$\pm$0.31 & 0.87$\pm$0.32 &  0.93$\pm$0.35  & 0.125 $\pm$ 0.079& 0.125 $\pm$ 0.039 & 0.121 $\pm$ 0.039  \\    
 u   &  -7.82$\pm$1.38 & -7.95$\pm$1.11 & -8.76$\pm$1.27 &  0.69$\pm$0.43 & 0.65$\pm$0.39 &  0.62$\pm$0.41  & 0.087 $\pm$ 0.138& 0.082 $\pm$ 0.069 & 0.071 $\pm$ 0.069 \\   
 g   &  -7.17$\pm$0.55 & -7.38$\pm$0.47 & -8.21$\pm$0.54 &  0.32$\pm$0.27 & 0.32$\pm$0.25 &  0.31$\pm$0.27  & 0.044 $\pm$ 0.055& 0.043 $\pm$ 0.027 & 0.038 $\pm$ 0.027 \\   
 r   &  -7.76$\pm$0.46 & -8.01$\pm$0.42 & -8.92$\pm$0.51 &  0.28$\pm$0.25 & 0.29$\pm$0.24 &  0.28$\pm$0.26  & 0.046 $\pm$ 0.046& 0.046 $\pm$ 0.023 & 0.042 $\pm$ 0.023  \\   
 i   &  -8.15$\pm$0.54 & -8.38$\pm$0.47 & -9.36$\pm$0.59 &  0.31$\pm$0.27 & 0.30$\pm$0.25 &  0.30$\pm$0.28  & 0.048 $\pm$ 0.055& 0.046 $\pm$ 0.027 & 0.039 $\pm$ 0.027 \\   
 z   &  -8.58$\pm$0.69 & -8.81$\pm$0.60 & -9.85$\pm$0.74 &  0.37$\pm$0.31 & 0.34$\pm$0.28 &  0.36$\pm$0.32  & 0.045 $\pm$ 0.086& 0.043 $\pm$ 0.043 & 0.042 $\pm$ 0.043 \\   
 J   &  -8.93$\pm$0.51 & -9.12$\pm$0.57 & -9.97$\pm$0.67 &  0.38$\pm$0.22 & 0.43$\pm$0.23 &  0.43$\pm$0.25  & 0.048 $\pm$ 0.042& 0.047 $\pm$ 0.021 & 0.044 $\pm$ 0.021 \\   
 H   &  -8.99$\pm$0.56 & -9.22$\pm$0.67 & -9.99$\pm$0.61 &  0.39$\pm$0.23 & 0.46$\pm$0.26 &  0.41$\pm$0.24  & 0.046 $\pm$ 0.071& 0.051 $\pm$ 0.035 & 0.044 $\pm$ 0.035 \\   
 K   &  -9.38$\pm$0.54 & -9.61$\pm$0.63 &-10.46$\pm$0.68 &  0.41$\pm$0.22 & 0.47$\pm$0.25 &  0.45$\pm$0.25  & 0.044 $\pm$ 0.045& 0.049 $\pm$ 0.023 & 0.043 $\pm$ 0.023  \\   
3.6  &  -9.56$\pm$0.46 & -9.37$\pm$0.42 &-10.08$\pm$0.40 &  0.42$\pm$0.20 & 0.44$\pm$0.19 &  0.38$\pm$0.19  & 0.043 $\pm$ 0.038& 0.046 $\pm$ 0.019 & 0.037 $\pm$ 0.019 \\   
4.5  &  -9.65$\pm$0.53 & -9.44$\pm$0.43 &-10.16$\pm$0.41 &  0.45$\pm$0.21 & 0.46$\pm$0.20 &  0.40$\pm$0.19  & 0.047 $\pm$ 0.044& 0.048 $\pm$ 0.022 & 0.040 $\pm$ 0.022 \\    
\hline
\end{tabular}
\caption{The slope, scatter an tightness for the TFrs, based on $2.2h$ magnitudes 
Column (1): band;
Column (2)-Column (4): slopes of the TFrs based on $W_{50}^{i}$, $V_{max}$ and $V_{flat}$, measured in magnitudes;
Column (5)-Column(7): scatter of the TFrs based on $W_{50}^{i}$, $V_{max}$ and $V_{flat}$, measured in magnitudes;
Column (8)-Column (10): tightness of the TFrs based on $W_{50}^{i}$, $V_{max}$ and $V_{flat}$, measured in dex;
}
\label{tbl_22h}
\end{table*}  
In this Appendix we briefly present the results of the TFrs, based on magnitudes measured within $2.2$ disk scale lengths.
Figure \ref{fig_sl22}  demonstrates the slope, scatter and tightness of these TFrs. 
It is clear from the figures, that even though the trends remain the same, the errors on the scatter and tightness
significantly increase. Moreover, usage of magnitudes measured within 2.2 disk scale lengths did not decrease the scatter or improve the 
tightness of the TFrs in
comparison with total magnitudes. Table \ref{tbl_22h} summarises the measurements for the TFrs, based on $2.2h$ magnitudes.

\end{document}